\documentclass{elsarticle}
\usepackage{amssymb}
\usepackage{multicol}
\usepackage{xcolor}
\usepackage{float}
\usepackage[titletoc]{appendix}
\usepackage{amsmath}
\usepackage{graphics}
\usepackage[T1]{fontenc}
\usepackage{rotate}
\usepackage{color}
\usepackage{wrapfig}
\usepackage{latexsym,epsfig,fancyheadings}
\usepackage{latexsym,epsfig}
\graphicspath{{Figures/}}
\usepackage[english,francais]{babel}
\DeclareMathOperator{\sign}{sgn}
\usepackage[margin=2.5cm]{geometry}

\def\og{\leavevmode\raise.3ex\hbox{$\scriptscriptstyle\langle\!\langle$~}}
\def\fg{\leavevmode\raise.3ex\hbox{~$\!\scriptscriptstyle\,\rangle\!\rangle$}}

\begin{document}

\begin{frontmatter}

\selectlanguage{english}
\title{New estimations of the added mass and damping of two cylinders vibrating in a viscous fluid, from theoretical and numerical approaches}

\selectlanguage{english}
\author[Affil1]{Romain Lagrange}
\address[Affil1]{Den-Service d'Etudes M\'ecaniques et Thermiques (SEMT), CEA, Universit\'e Paris-Saclay, F-91191, Gif-sur-Yvette, France}
\ead{romain.lagrange@cea.fr}
\ead{romain.g.lagrange@gmail.com}
%\author[Affil1]{Xavier Delaune}
%\ead{meunier@irphe.univ-mrs.fr}
%\author[Affil1]{Philippe Piteau}
%\ead{eloy@irphe.univ-mrs.fr}
%\author[Affil1]{Laurent Borsoi}
%\ead{francois.nadal@cea.fr}
\author[Affil2]{Yann Fraigneau}
\address[Affil2]{LIMSI, CNRS, Universit\'e Paris-Sud, Orsay 91405, France}

%\address[Affil2]{Centro de Ci\^encias e Tecnologias Nucleares, Instituto Superior T\'ecnico, Universidade de Lisboa, Estrada Nacional 10, Km 139.7, 2695-066 Bobadela LRS, Portugal}

% if you know the dates of reception, and acceptation you can put them now;
% idem for the name of the person presenting the Note

%\medskip
%\begin{center}
%{\small Received *****; accepted after revision +++++\\
%Presented by *****}
%\end{center}

\begin{abstract}

{This paper deals with the small oscillations of two circular cylinders immersed in a viscous stagnant fluid. 
A new theoretical approach based on an Helmholtz expansion and a bipolar coordinate system is presented to estimate the fluid forces acting on the two bodies. We show that these forces are linear combinations of the {\textcolor{black}{cylinder accelerations}} and velocities, through viscous fluid added coefficients. {\textcolor{black}{To assess the validity of this theory, we consider the case of two equal size cylinders, one of them being stationary while the other one is forced sinusoidally}}. The self-added mass and damping coefficients are shown to decrease with both the Stokes number and the separation distance. The cross-added mass and damping coefficients tend to increase with the Stokes number and the separation distance. Compared to the inviscid results, the effect of viscosity is to add a correction term which scales as $Sk^{-1/2}$. When the separation distance is sufficiently large, the two cylinders behave as if they were independent and the Stokes predictions for an isolated cylinder are recovered. Compared to previous works, the present theory offers a simple and flexible alternative for an easy determination of the fluid forces and related added coefficients. To our knowledge, this is also the first time that a numerical approach based on a penalization method is presented in the context of fluid-structure interactions for relatively small Stokes numbers, and successfully compared to theoretical predictions.} 

%{\it To cite this article: R. Lagrange, ? (2016).}

%\vskip 0.5\baselineskip
 
\end{abstract}


\end{frontmatter}

\selectlanguage{english}

\section{Introduction}

The determination of the fluid force acting on an immersed body has been the topic of considerable experimental and theoretical studies, covering a full range of applications, from turbomachinery \cite{Furber1979}, heat exchangers tube banks \cite{Chen1975b,Chen1977} to biomechanics of plants \cite{DeLangre2008} or energy harvesting of flexible structures \cite{Doare2011,Singh2012,Michelin2013,Virot2016,Eloy2008}. 
Early researches were stimulated by the need of understanding the effect of the inertia of a surrounding fluid on the frequency of an oscillating pendulum \cite{DuBuat1786}. Assuming an inviscid fluid, \cite{Poisson1832,Green1833,Stokes1843} showed that the {\textcolor{black}{fluid makes}} the mass of the pendulum to increase by a factor that depends on the fluid density and the geometry of the pendulum. Since these pioneer works, this apparent increase of mass has commonly been referred as the added mass concept. It has been investigated in various experiments \cite{1,3,4,6,12,13,17,20,29} in which a single body is accelerated in a fluid initially at rest. The acceleration of the body induces a fluid motion which in returns induces an inertia effect from which an added mass coefficient is computed. 

The concept of added mass also applies to multiple immersed bodies, although its formulation is more complex as it involves "self-added" and "cross-added" mass coefficients.
The self-added mass coefficient characterizes the force on a body due to its own motion. The cross-added coefficient characterizes the fluid-coupling force on a stationary body due to the motion of an other body. 
Considering multiple arrangements, many experimental rigs have been built \cite{10,18,19,26,27,30,31,Chen1975b,34,36,42} to obtain precise measurements of these coefficients. 
From a theoretical {\textcolor{black}{standpoint}}, the added coefficients should be computed from the Navier-Stokes equations. However, in many practical situations, the effects of fluid viscosity and compressibility are neglected and a potential theory is carried out. A method of images \cite{Hicks1879,Greenhill1882,Basset1888,Carpenter1958,Birkhoff1960,Gibert1980,Landweber1991} or a complex analyis based on conformal transformations \cite{Wang2004,Burton2004,Tchieu2010,Scolan2008,Crowdy2006,Crowdy2010} are usually derived to solve the boundary value problem governing the fluid potential function. 
For small amplitude motions not entailing flow separation, the potential theory will accurately give
the added mass coefficients, and tabulated results are
available in the literature for a wide variety of
immersed geometries \cite{Patton1965}. 

All of the above-mentioned studies have dealt with an ideal fluid, whereas the viscous effects may be important for some applications such as bodies relatively close to each other. 
Considering the small oscillations of a single body in a viscous fluid, Stokes \cite{Stokes1850} solved the linearized Navier-Stokes equations and showed that the fluid force is a linear combination of two components related to the acceleration of the body and its velocity. The coefficients of this linear combination are commonly {\textcolor{black}{referred to as}} the viscous added mass and the viscous added damping, respectively. Stokes found that the effect of viscosity is to add to the ideal fluid added mass coefficient a correction term which depends on the fluid mass density and viscosity, the frequency of oscillation, and a characteristic length scale. All of these effects can be regrouped in a single dimensionless number, the Stokes number. 
     
The extension of this work to the case of multiple bodies remains a challenging theoretical problem, mainly due to the viscous boundary conditions to account for. One approach developped in \cite{Chen1975b} is to associate to each body a fluid potential and a stream function, governed by a Laplace and an Helmholtz equation, respectively. Introducing a polar coordinate system attached to each body, a method of separation of variables is used to expand the {\textcolor{black}{potential and stream functions}} as an infinite trigonometric series with unknown coefficients. Applying the viscous boundary conditions into each local coordinate system yields a set of linear equations for these unknowns. The number of equations depends on the number of bodies and the number of terms used in the series expansions. In the end, the set of linear equations has to be solved numerically. The two cylinders problem could be solved in this framework, but even for such a restricted number of bodies, the method of \cite{Chen1975b} is hardly tractable. 

In this paper, we build on our previous work which dealt with ideal fluids \cite{Lagrange2018} to introduce a flexible theoretical method and obtain an estimation of the viscous added coefficients. 
In addition to this theoretical work, we perform some numerical simulations where the immersed boundary conditions are considered with a penalization method. The choice of this approach relies on its effectiveness and simplicity of implementation in CFD codes, without deep modification of the algorithmic structure. The basic idea is to add a forcing term in the Navier-Stokes equation set over the area of the immersed body in order to locally impose the velocity of the body \cite{peskin_2002}. The method does not require any mesh update related to the motion of the body, any complex geometrical considerations on the position of the wall in regard to the computational grid or any high order interpolations as done with some other approaches (e.g. ALE methods \cite{loubere_2010}, cut-cell methods \cite{cheny_2010}, immersed body methods \cite{gronski_2016}). In the present work, we actually use a variant method initially proposed by \cite{pasquetti_2008}, called the pseudo-penalization method, in which disappears the stiffness nature of the Navier-Stokes equations due to the forcing term. 
The penalization and pseudo-penalization methods are particularly efficient in fluid problems with moderate or high Reynolds numbers (see e.g.  \cite{mittal_2005,kadoch_2012,kolomenskiy_2011,schneider_2015,minguez_2008,nore_2018})
but has never been tested in problems with low Reynolds numbers, as considered in the present work.
\\
\\ 
This paper is organized as follows. Section \ref{Sec.DefinitionOfTheProblem} presents the problem and the governing equations for two circular cylinders immersed in
a viscous fluid at rest. In
{{\textcolor{black}{Section}} \ref{Sec_theory}, we propose a theoretical approach based on an Helmholtz decomposition and a bipolar coordinate system to obtain an approximate solution of the fluid problem. We derive expressions for the fluid potential and stream functions, from which we compute the fluid forces on the cylinders. In {{\textcolor{black}{Section}} \ref{Sec_numerics} we describe the numerical simulations that we have performed to solve the fluid problem. The results of our investigation are presented in {{\textcolor{black}{Section}} \ref{Sec_results}. Throughout, we directly compare the theoretical predictions to the numerical simulations. We start with comparing the time evolutions of the fluid forces acting on the cylinders, when one is stationary while the other is imposed a sinusoidal vibration. We
then analyze the dependance of the fluid added coefficients with the Stokes number and the separation distance. Some scaling laws are derived in the limit of large Stokes numbers. Finally, {{\textcolor{black}{Section}} \ref{Sec:Conclusion} summarizes our findings. 

%%%%%%%%%%%%%%%%%%%%%%%%%%%%%%%%%%%%%%%%%%%%%%%%%%%%%%%%%%%%%%%%%%%%%%
\begin{nomenclature} 
\begin{deflist}[AAAAAAAAA] %[AAAA] if you have 4 letters max for example 
\defitem{$O_j$}\defterm{center of cylinder ${\mathcal{C}_j}$} 
\defitem{$O$}\defterm{midpoint of $O_1$ and $O_2$} 
\defitem{$R_j$}\defterm{radius of cylinder ${\mathcal{C}_j}$} 
\defitem{$\Omega$}\defterm{angular frequency of the cylinders} 
\defitem{$T, t$}\defterm{dimensional and dimensionless time} 
\defitem{$\partial {C_j}$}\defterm{boundary of ${\mathcal{C}_j}$} 
\defitem{${{\bf{n}}_j}$}\defterm{outward normal unit vector to $\partial {C_j}$} 
\defitem{$E$}\defterm{separation distance}
\defitem{$\rho$}\defterm{fluid volume mass density} 
\defitem{$\nu$}\defterm{fluid kinematic viscosity} 
\defitem{${\bf{U}}_j$}\defterm{displacement vector of cylinder ${\mathcal{C}_j}$}
\defitem{$U$}\defterm{max of $\left(|{\bf{U}}_1|,|{\bf{U}}_2|\right)$}
\defitem{${\bf{{u}}}_j^*$}\defterm{dimensionless displacement vector of cylinder ${\mathcal{C}_j}$} 
\defitem{${\bf{{u}}}_j$}\defterm{complex dimensionless displacement vector of cylinder ${\mathcal{C}_j}$}  
\defitem{$u_{jx}, u_{jy}$}\defterm{$x$ and $y$ components of ${\bf{{u}}}_j$}
\defitem{${\bf{V}}, P$}\defterm{fluid flow velocity vector and pressure} 
\defitem{${\bf{{v}}}^*, {p}^*$}\defterm{dimensionless fluid flow velocity vector and pressure}
\defitem{${\bf{{v}}}, p$}\defterm{complex dimensionless fluid flow velocity vector and pressure} 
\defitem{${{{\bf{F}}_j}}$}\defterm{fluid force on cylinder ${\mathcal{C}_j}$}
\defitem{${\bf{{f}}}_j^*$}\defterm{dimensionless fluid force on cylinder ${\mathcal{C}_j}$}
 \defitem{${\bf{{f}}}_j$}\defterm{complex dimensionless fluid force on cylinder ${\mathcal{C}_j}$}
\defitem{$r$}\defterm{radius ratio}
\defitem{$\varepsilon$}\defterm{dimensionless separation distance} 
\defitem{$KC$}\defterm{Keulegan-Carpenter number} 
\defitem{$Sk$}\defterm{Stokes number}
\defitem{$\varphi, {\bf{A}}$}\defterm{fluid potential and stream functions} 
\defitem{$\widetilde{\varphi}, \widetilde{\bf{A}}$}\defterm{ad-hoc fluid potential and stream functions} 
\defitem{$\widetilde{\bf{{f}}}_j$}\defterm{ad-hoc fluid force on cylinder ${\mathcal{C}_j}$}
\defitem{$h_j, \phi_j$}\defterm{magnitude and phase angle of $\widetilde{\bf{{f}}}_j$}
\defitem{$z$}\defterm{complex cartesian coordinate} 
\defitem{$x, y$}\defterm{real and imaginary parts of $z$} 
\defitem{${\bf{e}}_x, {\bf{e}}_y$}\defterm{cartesian basis vectors}
\defitem{$\zeta$}\defterm{complex bipolar coordinate} 
\defitem{$\sigma, \tau$}\defterm{real and imaginary parts of $\zeta$} 
\defitem{${\bf{e}}_\sigma, {\bf{e}}_\tau$}\defterm{bipolar basis vectors}
\defitem{$\tau_j$}\defterm{bipolar coordinate of $\partial {C_j}$}
\defitem{$\kappa_{\sigma \tau }$}\defterm{Lamé coefficient of the bipolar coordinates system}
\defitem{$k$}\defterm{ad-hoc constant}
\defitem{$W$}\defterm{residual of the approximation}
\defitem{$k^{COL}, k^{LS}$}\defterm{ad-hoc constants for the collocation and least squares approximation methods}
\defitem{$[M],[C]$}\defterm{added mass and damping matrices}
\defitem{$m_{self}^{(j)}, c_{self}^{(j)}$}\defterm{self-added mass and damping coefficients}
\defitem{$m_{cross}$}\defterm{cross-added mass coefficient}
\defitem{$c_{cross}$}\defterm{cross-added damping coefficient}
\defitem{$m_{self}^{(j)POT}, m_{cross}^{POT}$}\defterm{inviscid limits of $m_{self}^{(j)}$ and $m_{cross}$}
\defitem{$m_{self}^{ISO}, c_{self}^{ISO}$}\defterm{self-added mass and damping coefficients of an isolated cylinder}
\defitem{$\delta t$}\defterm{time step of numerical simulations}
\defitem{$\chi$}\defterm{penalty function of numerical simulations}
\defitem{${\rm{K}}_j$}\defterm{modified Bessel function of second kind}
\defitem{$\iota$}\defterm{relative deviation between theoretical and numerical predictions}

\end{deflist} 
\end{nomenclature}

%%%%%%%%%%%%%%%%%%%%%%%%%%%%%%%%%%%%%%%%%%%%%%%%%%%%%%%%%%%%%%%%%%%%%%

\section{Definition of the problem and governing equations}\label{Sec.DefinitionOfTheProblem}

We consider the simple harmonic motions of two rigid circular cylinders ${\mathcal{C}_j}$, $\left(j = 1,2\right)$, with centers ${O_j}$, radii $R_j$, boundaries $\partial {C_j}$, immersed in an infinite 2D viscous fluid domain, as illustrated in {\textcolor{black}{Figure}} \ref{Fig1}. The angular frequency of the cylinders is $\Omega$ and their displacement vectors are ${\bf{U}}_j$.
The fluid is Newtonian, homogeneous, of volume mass density $\rho$ and kinematic viscosity $\nu$. The Navier-Stokes equations and the boundary conditions for the incompressible fluid flow $\left( {{\bf{V}} ,P} \right)$ write
\begin{subequations}\label{DimensionNavierStokes}
%\begin{align}
%\nabla  \cdot {\bf{V}}  &= 0,\\
%\frac{{\partial {\bf{V}} }}{{\partial T}} + \nabla {\bf{V}}  \cdot {\bf{V}}+ \frac{1}{\rho }\nabla P  - \nu \Delta {\bf{V}} &= {\bf{0}},\\
%{\bf{V}}  - \frac{d{\bf{U}}_j}{dT}&= {\bf{0}}\;\;\;\mbox{on $\partial {C_j}$}, j=\{1,2\}.
%%{\bf{V}}&\rightarrow{\bf{0}}\;\;\;\mbox{at infinity}.
%\end{align}
\begin{align}
\nabla  \cdot {\bf{V}}  &= 0,\\
\frac{{\partial {\bf{V}} }}{{\partial T}} + {\color{black}{{\left({\bf{V}} \cdot \nabla\right) {\bf{V}}}}} + \frac{1}{\rho }\nabla P  - \nu \Delta {\bf{V}} &= {\bf{0}},\\
{\bf{V}}  - \frac{d{\bf{U}}_j}{dT}&= {\bf{0}}\;\;\;\mbox{on $\partial {C_j}$}, j=\{1,2\}.
%{\bf{V}}&\rightarrow{\bf{0}}\;\;\;\mbox{at infinity}.
\end{align}

\end{subequations}
The third equation expresses the continuity of velocities at the cylinder boundaries. 
%The last equation states that the fluid is unperturbated far from the cylinders.
The fluid force acting on ${\mathcal{C}_j}$ is the sum of a pressure and a viscous term, and writes
\begin{align}\label{DimensionFluidForce}
{{{\bf{F}}_j } } &= - \int\limits_{\partial {C_j}} { {P } {\bf{n}}_j dL_j}  + \rho\nu\int\limits_{\partial {C_j}} {\left[ {\nabla  {{\bf{V}} }  + \left( {\nabla  {{\bf{V}}} } \right)^T } \right] \cdot {\bf{n}}_j dL_j }.
\end{align}
In this equation, ${{\bf{n}}_j}$ is the outward normal unit vector to $\partial {C_j}$, $\left( {\nabla  {{\bf{V}}} } \right)^T$ the transposate tensor of $ {\nabla  {{\bf{V}}} } $ and $d{L_j}$ an infinitesimal line element of integration. 

\begin{figure}[H]
\begin{center}
\includegraphics[width=1\textwidth]{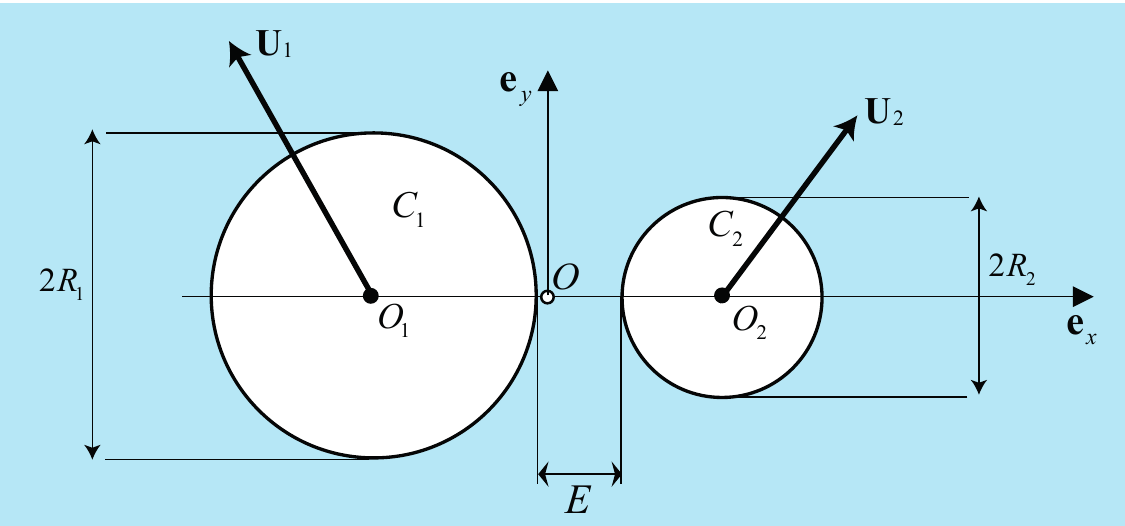}
\caption{Schematic diagram of the system: two oscillating cylinders ${\mathcal{C}_j}$ with radii ${R_j}$, centers ${O_j}$, displacement vectors ${{{\bf{U}}}}_j\left(T\right)$, are immersed in a fluid of kinematic viscosity $\nu$. The small oscillations of ${\mathcal{C}_j}$ generate an incompressible fluid flow. The midpoint of $O_1$ and $O_2$ is $O$ and the separation distance is $E$.}\label{Fig1}
\end{center}
\end{figure}

\subsection{Dimensionless equations}

In what follows, we use $R_2$ and $\Omega^{-1}$ as a characteristic length and time. Introducing $t = T\Omega$, we define the dimensionless {\textcolor{black}{cylinder displacements}} ${\bf{{u}}}_j^*$, fluid flow $\left({\bf{{v}}}^*,{p}^* \right)$ and fluid force ${\bf{{f}}}_j^*$  as 
\begin{equation}\label{DimensionlessParameters1}
{\bf{U}}_j = U \;{\bf{{u}}}_j^*, 
{\bf{V}} = U\Omega \;{\bf{{v}}}^*,
P= \rho U R_2 \Omega^2 \; {p}^*,
{\bf{F}}_j = {{\rho U\left( {R_2\Omega } \right)^2 }} \;{\bf{{f}}}_j^*,
\end{equation} 
with $U=\max\left(|{\bf{U}}_1|,|{\bf{U}}_2|\right)$.

To reduce the number of parameters of the problem we also introduce the rescaled quantities  
\begin{equation}\label{DimensionlessNumbers}
r = \frac{{{R_1}}}{{{R_2}}},\;\;\;\;{\varepsilon} = \frac{{{{{E}}}}}{{{R_2}}}, \;\;\;\; KC  = \frac{{U}}{{R_2 }},\;\;\;\; Sk  = \frac{{ {R_2}^2\Omega}}{{\nu }},
\end{equation}
as the radius ratio, separation distance, Keulegan-Carpenter number and Stokes number (i.e. vibration Reynolds number), respectively.

Introducing \eqref{DimensionlessParameters1} in \eqref{DimensionNavierStokes}, the dimensionless Navier-Stokes equations write 
\begin{subequations}\label{DimensionlessNavierStokes1}
\begin{align}
\nabla  \cdot {\bf{{v}}^*}  &= 0,\label{DimensionlessNavierStokes1a}\\
\frac{{\partial {\bf{{v}}^*} }}{{\partial t}} + KC{\color{black}{\left({\bf{{v}}^*} \cdot \nabla\right) {\bf{{v}}^*} }}+ \nabla {p^*}  - \frac{1}{Sk} \Delta {\bf{{v}}^*} &= {\bf{0}},\label{DimensionlessNavierStokes1b}\\
{\bf{{v}}^*}  - \frac{d{\bf{{u}}}_j^*}{dt}&= {\bf{0}}\;\;\;\mbox{on $\partial {C_j}$}, j=\{1,2\}.
%{\bf{V}}&\rightarrow{\bf{0}}\;\;\;\mbox{at infinity},
\end{align}
\end{subequations}
Introducing \eqref{DimensionlessParameters1} in \eqref{DimensionFluidForce}, the dimensionless fluid force acting on ${\mathcal{C}_j}$ write 
\begin{equation}\label{DimensionlessFluidForce}
 {{{\bf{{f}}}_j ^*} }  =  - \int\limits_{\partial {C_j}} { {{p}^* } {\bf{n}}_j dl_j}  + \frac{1}{{Sk}}\int\limits_{\partial {C_j}} {\left[ {\nabla  {{\bf{{v}}^*} }  + \left( {\nabla  {{\bf{{v}}^*}} } \right)^T } \right] \cdot {\bf{n}}_j dl_j }, 
\end{equation}
with $d{l_j}=d{L_j}/R_2$.

\section{Theoretical approach}\label{Sec_theory}

In the limit of small oscillations, i.e. $KC=o(1)$, the nonlinear convective term in the Navier-Stokes equations is negligible. Introducing ${\bf{{u}}}_j^* = \Re\{e^{{\rm{i}}t}{\bf{u}}_j\}$, ${\bf{{v}}^*} = \Re\{e^{{\rm{i}}t}{\bf{v}}\}$, ${p}^*= \Re\{e^{{\rm{i}}t} p\}$, the equations \eqref{DimensionlessNavierStokes1} rewrite
\begin{subequations}\label{DimensionlessNavierStokes}
\begin{align}
\nabla  \cdot {\bf{v}}  &= 0,\\
{\rm{i}}{\bf{v}} + \nabla p  - \frac{1}{Sk} \Delta {\bf{v}}  &= {\bf{0}},\\
{\bf{v}}  -{\rm{i}}{ {{\bf{u}}_j}}&= {\bf{0}}\;\;\;\mbox{on $\partial {C_j}$}, j=\{1,2\},
%{\bf{v}}&\rightarrow{\bf{0}}\;\;\;\mbox{at infinity},
\end{align}
\end{subequations}
with $\Re$ the real part operator \textcolor{black}{and ${\rm{i}}$ the imaginary unit.}

\subsection{Helmholtz decomposition}

We seek a solution of \eqref{DimensionlessNavierStokes} as a superposition of an irrotational and a divergence-free flow (Helmholtz decomposition)
\begin{equation}\label{HelmholtzDecomposition}
{{\bf{v}}}  = \nabla \varphi  + \nabla  \times {\bf{A}},
\end{equation}
with $\varphi$ and ${\bf{A}}=A{\bf{e}}_z$ some unknown potential and stream functions. Introducing this decomposition in \eqref{DimensionlessNavierStokes} yields 
\begin{subequations}\label{DimensionlessNavierStokes_Helmholtz}
\begin{align}
\Delta {\varphi}  &= 0,
\label{DimensionlessNavierStokes_Helmholtz_a}\\
\nabla  \times \left(  \Delta {\bf{A}}-{{\rm{i}}{Sk\bf{A}}} \right)-Sk\nabla \left( {{\rm{i}}\varphi  + p} \right) &= {\bf{0}},
\label{DimensionlessNavierStokes_Helmholtz_b}\\
\nabla \varphi+\nabla \times {\bf{A}} -{\rm{i}}{ {{\bf{u}}_j}}&= {\bf{0}}\;\;\;\mbox{on $\partial {C_j}$}, j=\{1,2\}.\label{DimensionlessNavierStokes_Helmholtz_c}
\end{align}
\end{subequations}
Taking the divergence and the curl of \eqref{DimensionlessNavierStokes_Helmholtz_b} yields two equations
\begin{equation}
p=-{\rm{i}}\varphi \;\;\; \mbox{and} \;\;\; \Delta A+\beta^2 A=0
\;\;\; \mbox{with} \;\;\; \beta=\sqrt{-{\rm{i}}Sk}, 
\label{Pressure}
\end{equation}
from which the pressure and the stream functions can be determined. 
 
\subsection{Bipolar coordinates}

Let $z=x+{\rm{i}}y$ be the complex number whose real and imaginary parts are the cartesian coordinates $x$ and $y$, measured from the midpoint $O$ of the two {\textcolor{black}{cylinder centers}}, $O_1$ and $O_2$, see {\textcolor{black}{Figure}} \ref{Fig_conformal_mappings}.

Let $h(z)$ be the conformal mapping defined as
\begin{equation}\label{Conformal_mapping_bipolar}
\zeta  = \sigma +{\rm{i}}\tau = h\left( z \right) = {\rm{i}}\ln \left( {\frac{{z - x_B  + a}}{{z - x_B  - a}}} \right),
\end{equation}
with $x_B= \left({r}^{2}-1\right)/\left(2d\right)$ and
\begin{equation}\label{KinematicsEquationsRepere}
a = \frac{{\sqrt {{d^2} - {{\left( {1 + r} \right)}^2}} \sqrt {{d^2} - {{\left( {1 - r} \right)}^2}} }}{2d}, 
\;\;\;\; d=r+\varepsilon+1.
\end{equation}
In \eqref{Conformal_mapping_bipolar}, $0<\sigma\leq2\pi$ and $\tau\in \mathbb{R}$ are the real and imaginary parts of $\zeta$, respectively. They are also the bipolar coordinates of a point in the plane $(x,y)$.
The images of ${\mathcal{C}_1}$ and ${\mathcal{C}_2}$ are the straight lines with ordinates ${\tau _1}$ and ${\tau _2}$ given by 
\begin{equation}\label{Tau_i}
{\tau _1} =  - {\sinh ^{ - 1}}\left( {{a \mathord{\left/
 {\vphantom {a r}} \right.
 \kern-\nulldelimiterspace} r}} \right)<0 \;\;\; \mbox{and} \;\;\; {\tau _2} = {\sinh ^{ - 1}}\left( a \right)>0.
\end{equation}

The Laplace operator and the fluid velocity vector in bipolar coordinates are
\begin{subequations}\label{Laplace_operator_bipolar}
\begin{align}
\Delta \varphi  &= {\left( {\frac{1}{{{\kappa _{\sigma \tau }}}}} \right)^2}\left( {\frac{{{\partial ^2}\varphi }}{{\partial {\sigma ^2}}} + \frac{{{\partial ^2}\varphi }}{{\partial {\tau ^2}}}} \right),
\\
{\bf{v}}  &= \frac{1}{{{\kappa _{\sigma \tau }}}}\left[\left( \frac{{\partial \varphi }}{{\partial \sigma }}+\frac{{{\partial }A }}{{\partial {\tau }}}\right){{\bf{e}}_\sigma } + \left(\frac{{\partial \varphi }}{{\partial \tau }}- \frac{{{\partial }A }}{{\partial {\sigma }}}\right){{\bf{e}}_\tau } \right],
\end{align}
\end{subequations}
with  $\kappa_{\sigma \tau }  = {a \mathord{\left/
 {\vphantom {a {\left[ {\cosh \left( \tau  \right) - \cos \left( \sigma  \right)} \right]}}} \right.
 \kern-\nulldelimiterspace} {\left[ {\cosh \left( \tau  \right) - \cos \left( \sigma  \right)} \right]}}
$ the Lam\'e coefficient and
\begin{equation}
{\bf{e}}_\sigma   = \frac{1}{\kappa_{\sigma \tau }} \left( {\frac{{\partial x}}{{\partial \sigma }}{\bf{e}}_x  + \frac{{\partial y}}{{\partial \sigma }}{\bf{e}}_y } \right),\;\;\;
{\bf{e}}_\tau   = \frac{1}{\kappa_{\sigma \tau }} \left( {\frac{{\partial x}}{{\partial \tau }}{\bf{e}}_x  + \frac{{\partial y}}{{\partial \tau }}{\bf{e}}_y } \right),
\end{equation}
the physical basis vectors. 
The fluid equations \eqref{DimensionlessNavierStokes_Helmholtz} in the bipolar coordinates system write
\begin{subequations}\label{Fluid_equations_bipolar}
\begin{align}
\frac{{{\partial ^2}\varphi }}{{\partial {\sigma ^2}}} + \frac{{{\partial ^2}\varphi }}{{\partial {\tau ^2}}} &= 0,
\label{Laplace_bipolar}
\\
 {\frac{{\partial ^2 A }}{{\partial \sigma ^2 }} + \frac{{\partial ^2 A }}{{\partial \tau ^2 }} + \beta ^2 {\kappa_{\sigma \tau }} ^2 A}  &= 0,
\label{Helmholtz_bipolar}
\\
\frac{{\partial \varphi }}{{\partial \sigma }}+\frac{{\partial A }}{{\partial \tau }} &= ({{ {\rm{i}} u}_{jx}}){g_{jy}} - ({{ {\rm{i}} u}_{jy}}){g_{jx}} \;\;\;\mbox{on $\tau  = {\tau_j},j=\{1,2\}$},
\label{BC1_bipolar}
\\
\frac{{\partial \varphi }}{{\partial \tau }}-\frac{{\partial A }}{{\partial \sigma }} &= ({{ {\rm{i}} u}_{jx}}){g_{jx}} + ({{ {\rm{i}} u}_{jy}}){g_{jy}} \;\;\;\mbox{on $\tau  = {\tau_j},j=\{1,2\}$},
\label{BC2_bipolar}
\end{align}
\end{subequations}
with ${g_{jx}} ={\kappa _{\sigma \tau_j }} {{{{\bf{e}}_x} \cdot {{\bf{e}}_{\tau_j} }}}$, ${g_{jy}} ={\kappa _{\sigma \tau_j }} {{{{\bf{e}}_y} \cdot {{\bf{e}}_{\tau_j} }}}$. These are $2\pi$ periodic functions of $\sigma$ given by 
\begin{subequations}\label{g_functions_bipolar}
\begin{align}
{g_{jx}}\left( \sigma  \right) & =  - a\frac{{\cos \left( \sigma  \right)\cosh \left( {{\tau _j}} \right) - 1}}{{{{\left( {\cosh \left( {{\tau_j}} \right) - \cos \left( \sigma  \right)} \right)}^2}}} = \sum\limits_{n = 1}^\infty  {{g_{jn}}\cos \left( {n\sigma } \right)} ,\\
{g_{jy}}\left( \sigma  \right) &=  - a\frac{{\sin \left( \sigma  \right)\sinh \left( {{\tau _j}} \right)}}{{{{\left( {  \cosh \left( {{\tau _j}} \right) - \cos \left( \sigma  \right)} \right)}^2}}} = \sum\limits_{n = 1}^\infty  {{g_{jn}} \sign\left(\tau_j\right)\sin \left( {n\sigma } \right)},
\end{align}
\end{subequations}
with ${g_{jn}} =  - 2na{e^{ -n \left|{\tau _j}\right|}}$.

\begin{figure}[H]
\begin{center}
\includegraphics[width=1\textwidth]{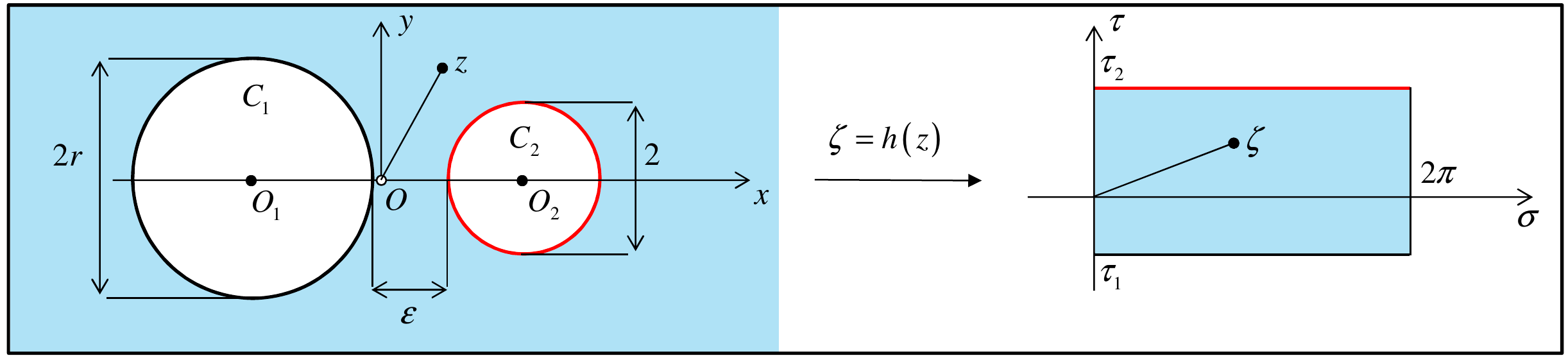}
\caption{Sketch of the conformal mapping $\zeta=h(z)$ defined by equation \eqref{Conformal_mapping_bipolar}.  The conformal function $\zeta=h(z)$ maps $\mathcal{C}_1$ and $\mathcal{C}_2$ into two parallel lines of equation $\zeta={\rm{i}}\tau_1$ and $\zeta={\rm{i}}\tau_2$.}\label{Fig_conformal_mappings}
\end{center}
\end{figure}

\subsection{Ad-hoc problem, fluid forces and added coefficients}

Since the problem is linear in ${ {u}}_{jx}$ and ${ {u}}_{jy}$, {\textcolor{black}{the functions $\varphi$ and $A$}} are linear combinations of the form 
\begin{subequations}
\begin{align}
\varphi & = \left( {u_{1x} \varphi _{1x}  + u_{2x} \varphi _{2x} } \right) + \left( {u_{1y} \varphi _{1y}  + u_{2y} \varphi _{2y} } \right),\\
A &= \left( {u_{1x} A_{1x}  + u_{2x} A_{2x} } \right) + \left( {u_{1y} A_{1y}  + u_{2y} A_{2y} } \right).
\end{align}
\end{subequations} 
The difficulty in finding $\varphi_{j\alpha}$ and $A_{j\alpha}$ arises from the fact that the Helmolhz equation \eqref{Helmholtz_bipolar} has a variable coefficient, $\kappa_{\sigma\tau}$. 
Instead, we consider the ad-hoc problem in which $\kappa_{\sigma\tau}$ is replaced by some unknown constant $k$, that will be determined later on. A method of separation of variables is then used to find the ad-hoc functions $\widetilde\varphi_{j\alpha}$ and $\widetilde A_{j\alpha}$. The boundary conditions \eqref{BC1_bipolar}, \eqref{BC2_bipolar} along with \eqref{g_functions_bipolar} indicate that $\widetilde\varphi_{j\alpha}$ and $\widetilde A_{j\alpha}$ are linear combinations of $\cos\left(n\sigma\right)$ and $\sin\left(n\sigma\right)$. Introducing these linear combinations in the Laplace and the Helmholtz equations, we also obtain that $\widetilde\varphi_{j\alpha}$ (resp. $\widetilde A_{j\alpha}$) is a linear combination of $\cosh\left(n\tau\right)$ and $\sinh\left(n\tau\right)$ (resp. $\cosh\left(l\tau\right)$ and $\sinh\left(l\tau\right)$ with $l=\sqrt{n^2-\left(\beta k\right)^2}$.\\
All in all, the ad-hoc functions write
\begin{subequations}\label{Fluid_functions_bipolar}  
\begin{align}
\widetilde\varphi  &= \left( {{\rm{i}}u_{1x} } \right)\sum\limits_{n = 1}^\infty  {\cos \left( {n\sigma } \right)\left[ {\varphi _n^{(1)} \left( {\tau _1 ,\tau _2 ,l } \right)\cosh \left( {n\tau } \right) + \varphi _n^{(2)} \left( {\tau _1 ,\tau _2 ,l } \right)\sinh \left( {n\tau } \right)} \right]}\nonumber
\\ 
&-\left( {{\rm{i}}u_{1y} } \right)\sum\limits_{n = 1}^\infty  {\sin \left( {n\sigma } \right)\left[ {\varphi _n^{(1)} \left( {\tau _1 ,\tau _2 ,l } \right)\cosh \left( {n\tau } \right) + \varphi _n^{(2)} \left( {\tau _1 ,\tau _2 ,l } \right)\sinh \left( {n\tau } \right)} \right]}\nonumber
\\
&+\left( {{\rm{i}}u_{2x} } \right)\sum\limits_{n = 1}^\infty  {\cos \left( {n\sigma } \right)\left[ {\varphi _n^{(1)} \left( {\tau _2 ,\tau _1 ,l } \right)\cosh \left( {n\tau } \right) + \varphi _n^{(2)} \left( {\tau _2 ,\tau _1 ,l } \right)\sinh \left( {n\tau } \right)} \right]}\nonumber
\\
&+\left( {{\rm{i}}u_{2y} } \right)\sum\limits_{n = 1}^\infty  {\sin \left( {n\sigma } \right)\left[ {\varphi _n^{(1)} \left( {\tau _2 ,\tau _1 ,l } \right)\cosh \left( {n\tau } \right) + \varphi _n^{(2)} \left( {\tau _2 ,\tau _1 ,l } \right)\sinh \left( {n\tau } \right)} \right]},\label{Potential_function_bipolar}
\\
\widetilde A  &= \left( {{\rm{i}}u_{1x} } \right)\sum\limits_{n = 1}^\infty  {\sin \left( {n\sigma } \right)\left[ {A _n^{(1)} \left( {\tau _1 ,\tau _2 ,l } \right)\cosh \left( {l\tau } \right) + A _n^{(2)} \left( {\tau _1 ,\tau _2 ,l } \right)\sinh \left( {l\tau } \right)} \right]}\nonumber
\\ 
&+\left( {{\rm{i}}u_{1y} } \right)\sum\limits_{n = 1}^\infty  {\cos \left( {n\sigma } \right)\left[ {A _n^{(1)} \left( {\tau _1 ,\tau _2 ,l } \right)\cosh \left( {l\tau } \right) + A _n^{(2)} \left( {\tau _1 ,\tau _2 ,l } \right)\sinh \left( {l\tau } \right)} \right]}\nonumber
\\
&+\left( {{\rm{i}}u_{2x} } \right)\sum\limits_{n = 1}^\infty  {\sin \left( {n\sigma } \right)\left[ {A _n^{(1)} \left( {\tau _2 ,\tau _1 ,l } \right)\cosh \left( {l\tau } \right) + A _n^{(2)} \left( {\tau _2 ,\tau _1 ,l } \right)\sinh \left( {l\tau } \right)} \right]}\nonumber
\\
&-\left( {{\rm{i}}u_{2y} } \right)\sum\limits_{n = 1}^\infty  {\cos \left( {n\sigma } \right)\left[ {A _n^{(1)} \left( {\tau _2 ,\tau _1 ,l } \right)\cosh \left( {l\tau } \right) + A _n^{(2)} \left( {\tau _2 ,\tau _1 ,l } \right)\sinh \left( {l\tau } \right)} \right]},\label{Stream_function_bipolar}
\end{align}
\end{subequations}  
with $\varphi _n^{(j)}$ and $A _n^{(j)}$ given in Appendix \ref{Appendix_Constants_Integration}. 

Plugging the Helmholtz decomposition $\widetilde{{\bf{v}}}  = \nabla \widetilde\varphi  + \nabla  \times \widetilde{\bf{A}}$ and the pressure equation $\widetilde p=-{\rm{i}}\widetilde\varphi$ {\textcolor{black}{given by \eqref{Pressure}}}} in \eqref{DimensionlessFluidForce} yields the ad-hoc fluid forces ${\bf{{\widetilde f}}}_j^* = \Re\{ e^{{\rm{i}}t}{\bf{\widetilde f}}_j\}$   
\begin{equation}\label{EqFluidForces}
\left( {\begin{array}{*{20}{c}}
{{{{\widetilde f}}_{1x}}}\\
{{{{\widetilde f}}_{1y}}}\\
{{{{\widetilde f}}_{2x}}}\\
{{{{\widetilde f}}_{2y}}}
\end{array}} \right) = \pi\left(\left[M\right] -\rm{i}\left[C\right]\right) \left( {\begin{array}{*{20}{c}}
{{{{u}}_{1x}}}\\
{{{{u}}_{1y}}}\\
{{{{u}}_{2x}}}\\
{{{{u}}_{2y}}}
\end{array}} \right),
\end{equation}
with $\left[M\right]$ and $\left[C\right]$ the added mass and damping matrices
\begin{equation}\label{AddedMatrices}
\left[M\right] = \left( {\begin{array}{*{20}{c}}
{{m_{self}^{(1)}}}&0&m_{cross} &0\\
0&{{m_{self}^{(1)}}}&0&{ - m_{cross} }\\
m_{cross} &0&{{m_{self}^{(2)}}}&0\\
0&{ - m_{cross} }&0&{{m_{self}^{(2)}}}
\end{array}} \right), 
\left[C\right] = \left( {\begin{array}{*{20}{c}}
{{c_{self}^{(1)}}}&0&c_{cross} &0\\
0&{{c_{self}^{(1)}}}&0&{ - c_{cross} }\\
c_{cross} &0&{{c_{self}^{(2)}}}&0\\
0&{ - c_{cross} }&0&{{c_{self}^{(2)}}}
\end{array}} \right).
\end{equation}
The self-added mass ${{m_{self}^{(j)}}}$ and damping ${{c_{self}^{(j)}}}$ relate the fluid force on $\mathcal{C}_j$ to its own motion. The cross-added mass $m_{cross}$ and damping $c_{cross}$ relate the fluid force on $\mathcal{C}_m$ to the motion of $\mathcal{C}_j$, $j\neq m$. 

All the fluid added coefficients in \eqref{AddedMatrices} are functions of the radius ratio $r$, the dimensionless separation distance $\varepsilon$ and the Stokes number $Sk$. A general closed-form expression for these coefficients is not tractable, but some simplifications are possible in particular cases. For example, as $Sk\rightarrow\infty$ (inviscid fluid), the flow is purely potential, i.e. $(\widetilde A, {{c_{self}^{(j)}}}, c_{cross})\rightarrow \left(0,0,0\right)$, and the added mass coefficients simplify to
\begin{subequations}\label{EqAddedMass_SkInf}
\begin{align}
m_{self}^{(1)}  &\rightarrow m_{self}^{(1)POT}= \sum\limits_{n = 1}^\infty  {\frac{{4 n{a^2}{e^{  2n{\tau _1}}}}}{{\tanh \left[ {n\left( {{\tau _2} - {\tau _1}} \right)} \right]}}}\;\; {\rm{as}}\;\; Sk\rightarrow\infty,\\
m_{self}^{(2)}  &\rightarrow m_{self}^{(2) POT}= \sum\limits_{n = 1}^\infty  {\frac{{4 n{a^2}{e^{ - 2n{\tau _2}}}}}{{\tanh \left[ {n\left( {{\tau _2} - {\tau _1}} \right)} \right]}}}\;\; {\rm{as}}\;\; Sk\rightarrow\infty,\\
m_{cross}   &\rightarrow  m_{cross}^{POT}= \sum\limits_{n = 1}^\infty  {\frac{{ - 4 n{a^2}{e^{ - n\left( {{\tau _2} - {\tau _1}} \right)}}}}{{\sinh \left[ {n\left( {{\tau _2} - {\tau _1}} \right)} \right]}}}\;\; {\rm{as}}\;\; Sk\rightarrow\infty. 
\end{align}
\end{subequations}
For the sake of clarity, we have reported the study of the variations of $m_{self}^{(j)POT}$ and $m_{cross}^{POT}$ in appendix \ref{Appendix_Inviscid}. 

\subsection{Determination of the ad-hoc constant $k$} 

In the previous section, we have obtained solutions of an ad-hoc problem in which the Lam\'e coefficient $\kappa_{\sigma\tau}$ has been replaced by some constant $k$. As a result, the ad-hoc functions $\widetilde A$, $\widetilde\varphi$ and $\widetilde p$ do not satisfy the Navier-Stokes equation \eqref{DimensionlessNavierStokes_Helmholtz_b}, leading to a non zero local residual
\begin{equation}
{\bf{W}}=u_{1x} {\bf{W}}_{1x}+u_{2x} {\bf{W}}_{2x} +u_{1y} {\bf{W}}_{1y}+u_{2y} {\bf{W}}_{2y},
\end{equation} 
with
${\bf{W}}_{j\alpha}=\nabla  \times \left(  \Delta {\bf{\widetilde A}}_{j\alpha}-{{\rm{i}}{Sk\bf{\widetilde A}}}_{j\alpha} \right)$ and ${\bf{\widetilde A}}_{j\alpha}={{\widetilde A}}_{j\alpha}{\bf{e}}_z.$
The constant $k$ is determined from the condition that the weigthed residual 
\begin{equation}\label{EqWeightedResidual}
W=\int\limits_0^{2\pi } {\int\limits_{\tau _1 }^{\tau _2 } \left({\left|{\bf{W}}_{1x}\right|w_{1x}+ \left|{\bf{W}}_{2x}\right|w_{2x} + \left|{\bf{W}}_{1y}\right|w_{1y}+ \left|{\bf{W}}_{2y}\right|w_{2y}}\right) } {\kappa_{\sigma \tau }} ^2  d\tau d\sigma,
\end{equation} 
must vanish for some given weight functions $w_{j\alpha}$. 
In this study, we consider two families of weight functions, {\textcolor{black}{which yield}} two sets of ad-hoc functions.    
In the least squares method, the weight functions are chosen in the form 
\begin{align} 
w_{j\alpha}  &= \frac{d}{{dk  }}\left|{\bf{W}}_{j\alpha}\right|,
\end{align}
such that the residual $W$ vanishes when
\begin{equation} 
\chi\left(k\right)= \int\limits_0^{2\pi } {\int\limits_{\tau _1 }^{\tau _2 } \left({\left|{\bf{W}}_{1x}\right|^2+ \left|{\bf{W}}_{2x}\right|^2 + \left|{\bf{W}}_{1y}\right|^2+ \left|{\bf{W}}_{2y}\right|^2}\right) } {\kappa_{\sigma \tau }} ^2  d\tau d\sigma,
\end{equation}
is minimum. We call $\chi^{LS}$ this minimum, reached for $k=k^{LS}$. 

In the collocation method, the residual $W$ is forced to vanish on the {\textcolor{black}{cylinder boundaries}}. The weight functions are chosen  to be the Dirac functions $\delta$
\begin{align} 
w_{j\alpha}  &= \frac{d}{{dk}}\left|{\bf{W}}_{j\alpha}\right|\left(\delta\left(\tau-\tau_1\right)+\delta\left(\tau-\tau_2\right)\right),
\end{align}
such that the residual $W$ vanishes when
\begin{align}
\chi\left(k\right)=&\int\limits_0^{2\pi } { {\left({\left|{\bf{W}}_{1x}\right|^2+ \left|{\bf{W}}_{2x}\right|^2 + \left|{\bf{W}}_{1y}\right|^2+ \left|{\bf{W}}_{2y}\right|^2}\right)}} {\kappa_{\sigma \tau }} ^2  \left( {\sigma ,\tau_1} \right) d\sigma\nonumber \\
+&\int\limits_0^{2\pi } { {\left({\left|{\bf{W}}_{1x}\right|^2+ \left|{\bf{W}}_{2x}\right|^2 + \left|{\bf{W}}_{1y}\right|^2+ \left|{\bf{W}}_{2y}\right|^2}\right)}} {\kappa_{\sigma \tau }} ^2  \left( {\sigma ,\tau_2} \right) d\sigma,
\end{align}
is minimum. We call $\chi^{COL}$ this minimum, reached for $k=k^{COL}$.
\\
\\
The evolutions of $k^{LS}$, $k^{COL}$, $\chi^{LS}$ and $\chi^{COL}$, versus the Stokes number $Sk$ are shown in Fig.~\ref{Fig2}, for equal size cylinders ($r=1$) and three dimensionless separation distances $\varepsilon=\{0.5, 1, 2\}$. We find that both $k^{LS}$ and $k^{COL}$ decrease with $Sk$, increase with $\varepsilon$, but remain close to $1$. This can be explained from the fact that the bipolar coordinates $(\sigma,\tau)$ are conformally equivalent to the cartesian coordinates $(x,y)$, in which the Helmholtz equation is similar to \eqref{Helmholtz_bipolar} under the change $(\sigma,\tau,\kappa_{\sigma\tau})\rightarrow(x,y,1)$.
The evolutions of $\chi^{LS}$ and $\chi^{COL}$ indicate that the theory becomes less accurate as the Stokes number and the dimensionless separation distance decrease (i.e. as the viscous and the confinement effects becomes preponderant).

\begin{figure}[H]
\begin{center}
\includegraphics[width=1\textwidth]{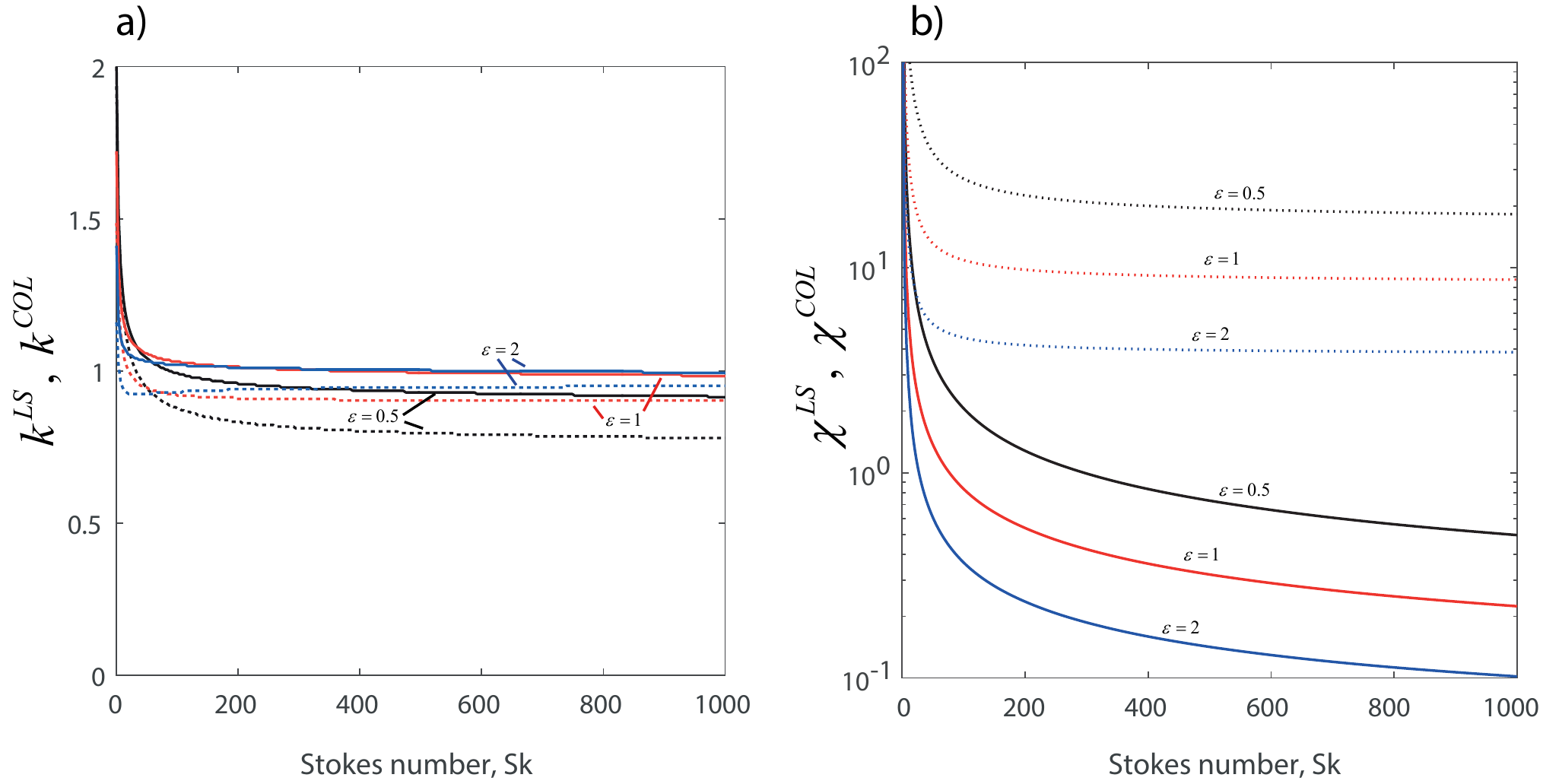}
\caption{Evolutions of $k^{LS}$, $k^{COL}$, $\chi^{LS}$, $\chi^{COL}$, versus the Stokes number $Sk$. The solid lines refer to the least squares approximation {\textcolor{black}{(LS)}} and the dotted lines refer to the collocation approximation {\textcolor{black}{(COL)}}. The dimensionless separation distance is $\varepsilon=0.5$ (black color), $\varepsilon=1$ (red color) and $\varepsilon=2$ (blue color). The radius ratio is $r=1$.}\label{Fig2}
\end{center}
\end{figure}

\section{Numerical simulation}\label{Sec_numerics}

\subsection{Solving the Navier-Stokes equations}\label{NumSolving}

The numerical method to solve the Navier-Stokes equations \eqref{DimensionlessNavierStokes1} is based on the projection method of \cite{guermond_2006} and the delta formulation of \cite{goda_1979}. The equations are discretized following a finite volume approach on a staggered structured grid (MAC procedure) with a second order approximation in time and space. A differentiation forumula (BDF2) is used for the time discretization of \eqref{DimensionlessNavierStokes1b}, leading to 
%The vectorial notation is cancelled for the sake of clarity. 
%The pressure is located at the cell center whereas the velocity components are placed at the center of cell faces.
\begin{equation}\label{disc_ns}
\frac{3{\bf{{v}}^*}^{(n+1)}}{2 \delta t}+\frac{-4{\bf{{v}}^*}^{(n)}+{\bf{{v}}^*}^{(n-1)}}{2 \delta t}+{\bf{NL}}^{(n+1)}  + \nabla {p^*}^{(n)} - \frac{1}{Sk} \Delta {\bf{{v}}^*}^{(n+1)}= {\bf{0}},
\end{equation}
with ${\bf{NL}}=KC\;\nabla {\bf{{v}}^*}\cdot {\bf{{v}}^*}$ and $n$ the subscript for the time step.
The convective term at time $(n+1)\delta t$ is computed from a linear extrapolation of the estimated values at time $n\delta t$ and $(n-1)\delta t$, i.e. ${\bf{NL}}^{(n+1)}= 2{\bf{NL}}^{(n)}-{\bf{NL}}^{(n-1)}$. The space discretization of the convective and viscous terms are approximated with a second order centered-scheme. An implicit discretization is applied to the viscous term in order to increase the numerical stability. The pressure gradient is explicitly defined, as suggested in the projection method.

Introducing $ \delta {v_i^*}^{(n+1)}={v_i^*}^{(n+1)}-{v_i^*}^{(n)}$ as the time increment of the $i$-th component of the velocity vector $\bf{{v}}^*$, the equation \eqref{disc_ns} reduces to a Helmholtz equation 
\begin{equation}\label{NumHelm}
 \delta {v_i^*}^{(n+1)}- \frac{2}{3}\frac{\delta t}{Sk}\Delta\left[\delta {v_i^*}^{(n+1)}\right]= S_i^{(n,n-1)}, 
\end{equation}
where $ S_i^{(n,n-1)}$ contains all the explicit terms of \eqref{disc_ns}.
{\textcolor{black}{Equation}} \eqref{NumHelm} is solved by means of an Alternating Direction Implicit method, see \cite{Peaceman1955}.

The Helmholtz decomposition of ${{\bf{{v}}^*}^{(n+1)}}$ with a potential function $\Phi$ yields the two equations
\begin{equation}\label{NumPressure}
\Delta \Phi = \frac {\nabla \cdot {{\bf{{v}}^*}^{(n+1)}}}{\delta t} \;\;\;\;\;\; {\rm{and}} \;\;\;\;\;\; \Phi={p^*}^{(n+1)}-{p^*}^{(n)} - \frac{1}{Sk} \nabla \cdot {{\bf{{v}}^*}^{(n+1)}}.
\end{equation}
The Poisson's equation is solved using a direct method based on the partial diagonalization of the Laplace operator. Having determined $\Phi$, the pressure at time $(n+1)\delta t$ is computed from the second equation of \eqref{NumPressure}. Finally, the velocity field ${\bf{{v}}^*}^{(n+1)}$ is corrected in order to satisfy the divergence-free condition 
\begin{equation}\label{upd_vel}
    {\bf{{v}}^*}^{(n+1)}:= {\bf{{v}}^*}^{(n+1)}-\frac{3}{2}\delta t \nabla \Phi.
\end{equation}

\subsection{The pseudo penalization method}

The pseudo penalization method is based on the standard volume penalty method, see \cite{peskin_2002,mittal_2005,kadoch_2012}, and has shown to be effective in solving fluid-structure interaction problems involving moving bodies, see \cite{pasquetti_2008,nore_2018}. The principle is to solve some penalized Navier-Stokes equations over a single domain, instead of considering two separate domains (fluid and solid) interacting through a set of boundary conditions. The original contribution of \cite{pasquetti_2008} relies on the removal of specific terms in the Navier-Stokes equations in order to turn them into steady penalized Stokes equations in the solid domains, where the penalty term is directly provided by the time-discretization scheme. 

The penalization of \eqref{disc_ns} writes
\begin {align} \label{ns_penal}
 \frac{3{{\bf{v}}^*}^{(n+1)}}{2 \delta t} + \left(1-\chi\right)\left[\frac{-4{{\bf{v}}^*}^{(n)}+{{\bf{v}}^*}^{(n-1)}}{2 \delta t}+{\bf{NL}}^{(n+1)}\right] +\nabla {p^*}^{(n)} &- \frac{1}{Sk} \Delta {{\bf{v}}^*}^{(n+1)}\nonumber\\&={\bf{0}},
\end {align}
with $\chi$ a penalty function defined as $\chi= 1$ in the solid domains and $\chi=0$ in the fluid domain. In \eqref{ns_penal}, ${3{{\bf{v}}^*}^{(n+1)}}/(2 \delta t)$ can be seen as a forcing term that makes ${{\bf{v}}^*}$ to tend to zero in the solid domains. Although ${{\bf{v}}^*}$ does not strictly vanishes in the solid domains, the consistency of the method scales as $\sqrt{\delta t/Sk}$. Since the forcing term is provided by the time step, $3/(2 \delta t)$, it does not affect the stiffness of the equations, preventing spurious effects or stability constraints, unlike the standard penalization methods.

For a body moving with a velocity ${{\bf{v_0}}^*}$, \eqref{ns_penal} can be reformulated as
\begin {align}\label{PenalWithVelo}
 \frac{3{{\bf{v}}^*}^{(n+1)}}{2 \delta t} + \left(1-\chi\right)\left[\frac{-4{{\bf{v}}^*}^{(n)}+{{\bf{v}}^*}^{(n-1)}}{2 \delta t}+{\bf{NL}}^{(n+1)}\right] +\nabla {p^*}^{(n)} &- \frac{1}{Sk} \Delta {{\bf{v}}^*}^{(n+1)}\nonumber\\&= \chi \frac{3{{\bf{v_0}}^*}}{2 \delta t},
\end {align}
and solved with the numerical method {\textcolor{black}{mentioned}} in {{\textcolor{black}{Section}} \ref{NumSolving}.

\section{{\textcolor{black}{Presentation of a case study}}}

We now present the results of our predictions, considering the case in which ${\mathcal{C}}_1$ is stationary while ${\mathcal{C}}_2$ is imposed a sinusoidal displacement in the $x$ - direction. 
For the geometric parameters, we have investigated the case of two equal size cylinders, corresponding to a radius ratio $r=1$. Three representative values were chosen for the dimensionless separation distance (depicted in the insets of {\textcolor{black}{Figures}} \ref{Fig_Forces_eps_05}, \ref{Fig_Forces_eps_1} and \ref{Fig_Forces_eps_2}): a small gap, $\varepsilon=0.5$; a gap with size one radius, $\varepsilon=1$; and a large gap, $\varepsilon=2$.  
In the presentation of our results, we first consider the effect of the Stokes number $100\leq Sk\leq 900$ and the dimensionless separation distance on the time evolution of the fluid forces. We then analyze the evolution of the magnitude $h_j$ and phase $\phi_j$ of the forces, including the case $\varepsilon\rightarrow\infty$ for which Stokes \cite{Stokes1850} obtained
\begin{equation}\label{Equations_Stokes_coeff}
f_{2x}  = \pi \left( {m_{self}^{ISO}  - {\rm{i}}c_{self}^{ISO} } \right)u_{2x}  = \pi \left[ {1 + \frac{4}{{\sqrt {{\rm{i}}Sk} }}\frac{{{\mathop{\rm K}\nolimits} _1 \left( {\sqrt {{\rm{i}}Sk} } \right)}}{{{\mathop{\rm K}\nolimits} _0 \left( {\sqrt {{\rm{i}}Sk} } \right)}}} \right]u_{2x}, 
\end{equation}
with ${\rm{K}}_0$ and ${\rm{K}}_1$ the modified Bessel functions of second kind. 
We finally study the evolution of the fluid added coefficients and derive some scaling laws for large Stokes numbers. Throughout the study, we perform some numerical simulations to corroborate the theoretical predictions, also providing a discussion on the limitations of both approaches.  
%\subsection{Theoretical predictions and numerical setup}

\subsection{Theoretical predictions}

Since the problem is {\textcolor{black}{symmetric}} about the axis $\tau=0$, we have $\tau_1=-\tau_2$, $m_{self}^{(1)}=m_{self}^{(2)}=m_{self}$, $c_{self}^{(1)}=c_{self}^{(2)}=c_{self}$ and $m_{self}^{(1)POT}=m_{self}^{(2)POT}=m_{self}^{POT}$. 
The dimensionless ad-hoc fluid forces are computed from \eqref{EqFluidForces}, with ${\bf{u}}_{1}=\bf{0}$, $u_{2x}=-{\rm{i}}$ and $u_{2y}=0$, leading to
\begin{subequations}\label{Fluid_forces_complex_form}
\begin{align}
\widetilde f_{1x} &= \pi \left( {m_{cross}  - {\rm{i}}c_{cross} } \right)u_{2x}  =h_1 e^{{\rm{i}}\phi_1}u_{2x},\\
\widetilde f_{2x} & = \pi \left( {m_{self}  - {\rm{i}}c_{self} } \right)u_{2x}  = h_2 e^{{\rm{i}}\phi_2}u_{2x}.
\end{align}
\end{subequations}

\subsection{Numerical setup}

A study of the domain-, grid- and time-step independence studies is reported in

Concerning the numerical simulations, the computational domain size $L_x \times L_y$ is considered sufficiently large to minimize the end effects. For the small and medium separation distances ($\varepsilon=0.5$ and $\varepsilon=1$), we set $L_x \times L_y=20\times 17$. For $\varepsilon=2$, we set $L_x \times L_y=22\times 17$ so that the distance between the cylinders and the domain ends is similar to the cases $\varepsilon=0.5$ and $\varepsilon=1$. For all the simulations, the Keulegan-Carpenter number is set to $KC=10^{-2}$.

The cartesian grid is built with a regular distribution over the cylinder domains, including the displacement zone. The dimensionless cell size is $2\times 10^{-3}$ in both the $x$ and $y$ directions. It follows that the smallest spatial scale of our problem, i.e. the cylinder displacement, is discretized over ten square cells, which yields a satisfying spatial resolution.
The cell-size distribution outside the cylinder domain is performed with a hyperbolic tangent function and vary from $2\times 10^{-3}$ to $3.25 \times 10^{-2}$, with a maximum size ratio of $1.42 \%$. The mesh size is $3060 \times 1850$ for $\varepsilon=0.5$ and $\varepsilon=1$, and $3300 \times 1850$ for $\varepsilon=2$. The time step is set to $\delta t= 2 \times 10^{-3}$ for $Sk=100$ and $\delta t= 5 \times 10^{-3}$ for $Sk>100$.
Regarding the boundary conditions at the domain ends, the normal velocity is set to zero to ensure a null flow rate far from the cylinders and the normal derivative of the tangential component is imposed to zero. The normal component of the pressure gradient is also set to zero, which is the usual boundary condition for the pressure field when the flow rate is imposed. 
\\
{\textcolor{black}{When ${\mathcal{C}}_1$ is stationary and ${\mathcal{C}}_2$ is imposed a sinusoidal displacement in the $x$ - direction, the real dimensionless fluid forces write
\begin{subequations}
\begin{align}
f_{1x}^*\left( t \right) &= {m_{cross}}\sin \left( t \right) - {c_{cross}}\cos \left( t \right),\\
f_{2x}^*\left( t \right) &= {m_{self}}\sin \left( t \right) - {c_{self}}\cos \left( t \right).
\end{align}
\end{subequations}
To extract the added coefficients from the numerical simulations of the fluid forces, we introduce the Fourier inner product
\begin{equation}
\left\langle {f\left( t \right),g\left( t \right)} \right\rangle  = \frac{1}{\pi }\int\limits_0^{2\pi } {f\left( t \right)g\left( t \right)dt},
\end{equation}
and compute ${m_{self}}$, ${c_{self}}$, ${m_{cross}}$ and ${c_{cross}}$ from 
\begin{subequations}
\begin{align}
{m_{self}} = \frac{{\left\langle {f_{2x}^*\left( t \right),\sin \left( t \right)} \right\rangle }}{\pi }\;\;\;\;&{\rm{   and   }}\;\;\;\;{m_{cross}} = \frac{{\left\langle {f_{1x}^*\left( t \right),\sin \left( t \right)} \right\rangle }}{\pi },\\
{c_{self}} =  - \frac{{\left\langle {f_{2x}^*\left( t \right),\cos \left( t \right)} \right\rangle }}{\pi }\;\;\;\;&{\rm{   and   }}\;\;\;\;{c_{cross}} =  - \frac{{\left\langle {f_{1x}^*\left( t \right),\cos \left( t \right)} \right\rangle }}{\pi }.  
\end{align}
\end{subequations}
}}
{\textcolor{black}{
Finally, we shall note that a mesh size, time step and computational domain size independence study has been performed, see \ref{Appendix_Effect_mesh_time}. In this appendix, we clearly show that refining the mesh size, reducing the time step or increasing the computational domain size has no significant effect on the fluid coefficients predicted numerically. The parameters used in this study are therefore appropriately chosen to ensure the numerical convergence of our results.   
}}
\section{Results}\label{Sec_results}

%\begin {tabular} {|*{4}{c |}}
%\hline
%$\varepsilon$   &    0.5     & 1 &   2 \\
%\hline
%Domain size & $\left[20 , 17 \right]$ & $\left[20 , 17 \right]$ & $\left[22 , 17 \right]$ \\
%\hline
%Mesh size    & $\left[3060 \times 1850 \right]$  & $\left[3060 \times 1850 \right]$  & $\left[3300 \times 1850 \right]$  \\
%\hline
%\end  {tabular}

\subsection{Fluid forces}

{\textcolor{black}{The time evolutions of the fluid forces are depicted in {\textcolor{black}{Figures}} \ref{Fig_Forces_eps_05}, \ref{Fig_Forces_eps_1} and \ref{Fig_Forces_eps_2}. The theoretical predictions show that the forces are sinusoidal functions whose amplitude and phase depend on $Sk$ (viscous effects) and $\varepsilon$ (confinement effects). We observe that the amplitude of the fluid forces decreases with $Sk$ and $\varepsilon$, and is maximum for the moving cylinder.}} To study this sensitivity in more detail, we plot in {\textcolor{black}{Figure}} \ref{Gain_phase} {\textcolor{black}{a)}} the evolutions of the magnitude $h_j$ and the phase $\phi_j$. We observe that $h_j$ is maximum for the moving cylinder, diverges to infinity when $Sk\rightarrow 0$ and decreases to $h_1\rightarrow \pi |m_{cross}^{POT}|$ and $h_2\rightarrow \pi |m_{self}^{POT}|$ as $Sk\rightarrow \infty$ (inviscid fluid). The magnitude is also shown to be maximum for the small values of $\varepsilon$ (strong confinement) and to decrease to $h_1\rightarrow 0$ and $h_2\rightarrow \pi |m_{self}^{ISO}|$ as $\varepsilon\rightarrow\infty$ (isolated cylinders). Thus, as one would expect, the fluid forces are all the more intense as both the viscous and confinement effects are important.
\\
The {\textcolor{black}{Figure}} \ref{Gain_phase} b) shows that the forces are in phase opposition, i.e. $\phi_1=\phi_2+\pi$, with $\phi_1$ increasing from $\phi_1\rightarrow \pi/2$ as $Sk\rightarrow 0$ to $\phi_1\rightarrow \pi$ as $Sk \rightarrow \infty$. We note that the confinement has a very weak effect on the phase, leading to a slight increase of $\phi_j$ with $\varepsilon$. 
The variations of $\phi_j$ imply that the direction of the fluid forces depends on $Sk$ and, to a lesser extent on $\varepsilon$. From \eqref{Fluid_forces_complex_form}, the fluid forces vanish and reverse their direction when $\Re\{e^{{\rm{i}}t} \widetilde f_{jx}\}=\Re\{e^{{\rm{i}}(t+\phi_j)} h_j u_{jx}\}=h_j\sin\left(t+\phi_j\right)=0$, i.e. $t=-\phi_j+k\pi,k\in\mathbb{Z}$. At that time, the dimensionless displacement $u_2=\sin(t)$ of the moving cylinder equals $u_{2}^{*}=\pm\sin\left(\phi_j\right)$. In {\textcolor{black}{Figure}} \ref{Gain_phase} c), we show that the fluid forces cause the cylinders to attract (resp. repel) each other when $ - 1\le u_2<- \left| {u_2^*} \right|$ (resp. $ \left| {u_2^*} \right|\le u_2<1 $). In the narrow range $ - \left| {u_2^*} \right|\le u_2<\left| {u_2^*} \right|$, the cylinders are attracted (resp. repelled) to each other if the velocity of the moving cylinder is positive (resp. negative). 
An estimation of $u_2^*$ is made possible from the observation that it is weakly sensitive to $\varepsilon$  (at least for  $\varepsilon \ge 0.5 $) and thus can be approximated by its limit as  $\varepsilon\rightarrow\infty$. From \eqref{Equations_Stokes_coeff} and $u_2^*=\pm \sin(\phi_2)=\pm \sin(\arg(f_{2x}/u_{2x}))$, it comes that  
\begin{equation}
u_2^*\approx=\pm\sin\left(\arctan\left(\frac{c_{self}^{ISO}}{m_{self}^{ISO}}\right)\right)\approx\pm\frac{c_{self}^{ISO}}{\sqrt{\left(m_{self}^{ISO}\right)^2+\left(c_{self}^{ISO}\right)^2}},
\end{equation}
which is the equation of the green line ($\varepsilon\rightarrow\infty$) shown in {\textcolor{black}{Figure}} \ref{Gain_phase} c).
An asymptotic expansion of the modified Bessel functions ${\rm{K}}_j$ entering in the definition of $m_{self}^{ISO}$ and $c_{self}^{ISO}$, see \eqref{Equations_Stokes_coeff}, yields that ${u_2}^*=O\left( {Sk^{ - {1 \mathord{\left/
 {\vphantom {1 2}} \right.
 \kern-\nulldelimiterspace} 2}} } \right)$ as $Sk\rightarrow\infty$.

Finally, we note that the theoretical predictions for $h_j$ and $\phi_j$ are successfully corroborated by the numerical simulations, in the sense that similar trends are clearly recovered. Still, we note that the numerical simulations are poorly sensitive to $\varepsilon$ and slightly understimate the magnitude $h_2$ of the fluid force acting on the moving cylinder, especially in the range of low Stokes numbers. A detailed discussion on the differences between the theoretical and numerical approaches is reported in {{\textcolor{black}{Section}} \ref{sec:num_vs_theory}. 

\begin{figure}[H]
\begin{center}
\includegraphics[width=1\textwidth]{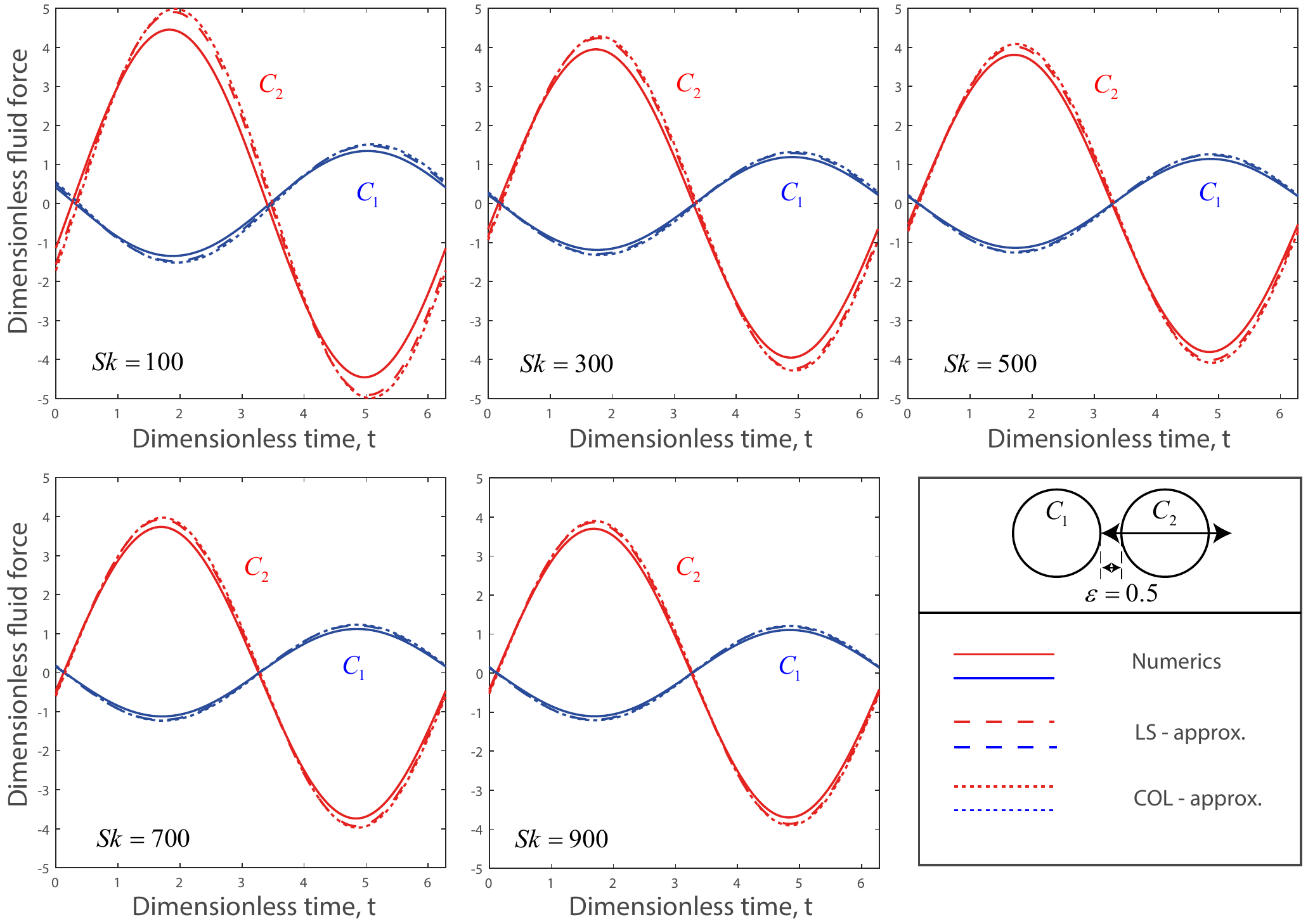}
\caption{Dimensionless fluid force $\Re\{e^{{\rm{i}}t} f_{jx}\}$ as a function of the dimensionless time $t$, for various Stokes numbers $Sk$. {\textcolor{black}{The dashed lines refer to the least squares approximation (LS) and the dotted lines refer to the collocation approximation (COL)}}.
The dimensionless separation distance is $\varepsilon=0.5$.}\label{Fig_Forces_eps_05}
\end{center}
\end{figure}

\begin{figure}[H]
\begin{center}
\includegraphics[width=1\textwidth]{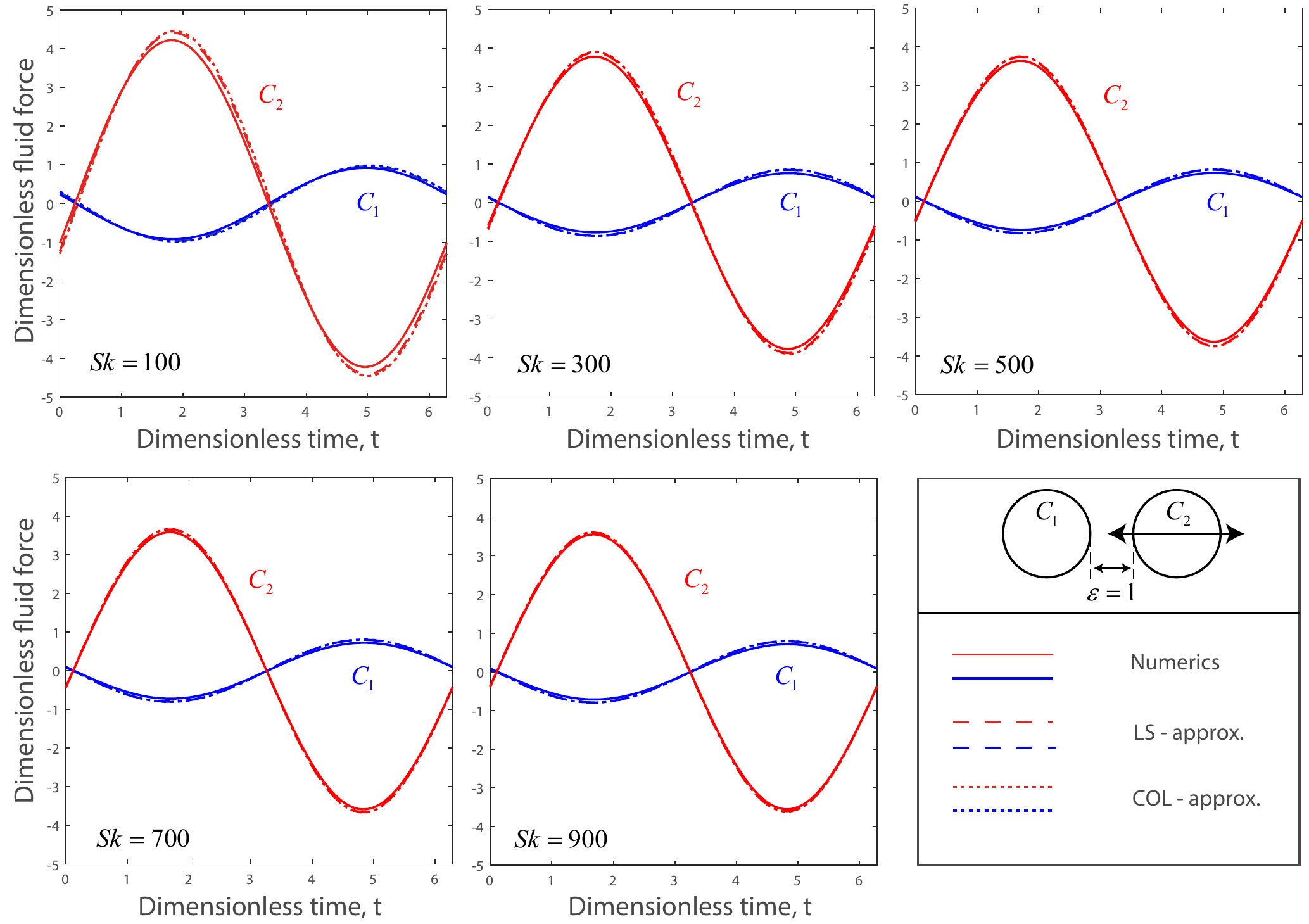}
\caption{Dimensionless fluid force $\Re\{e^{{\rm{i}}t} f_{jx}\}$ as a function of the dimensionless time $t$, for various Stokes numbers $Sk$. {\textcolor{black}{The dashed lines refer to the least squares approximation (LS) and the dotted lines refer to the collocation approximation (COL)}}. The dimensionless separation distance is $\varepsilon=1$.}\label{Fig_Forces_eps_1}
\end{center}
\end{figure}

\begin{figure}[H]
\begin{center}
\includegraphics[width=1\textwidth]{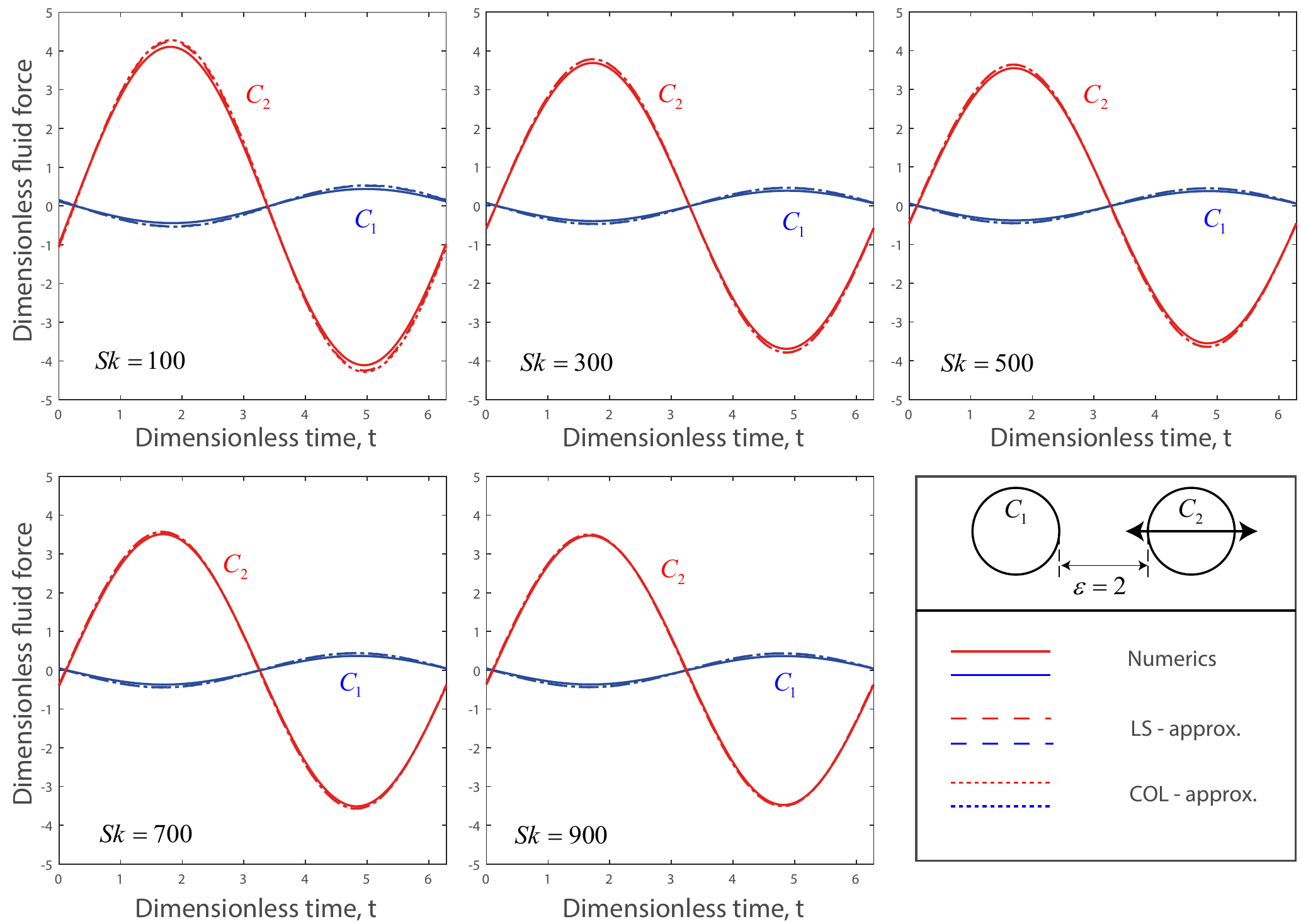}
\caption{Dimensionless fluid force $\Re\{e^{{\rm{i}}t} f_{jx}\}$ as a function of the dimensionless time $t$, for various Stokes numbers $Sk$. {\textcolor{black}{The dashed lines refer to the least squares approximation (LS) and the dotted lines refer to the collocation approximation (COL)}}. The dimensionless separation distance is $\varepsilon=2$.}\label{Fig_Forces_eps_2}
\end{center}
\end{figure}

\begin{figure}[H]
\begin{center}
\includegraphics[width=1\textwidth]{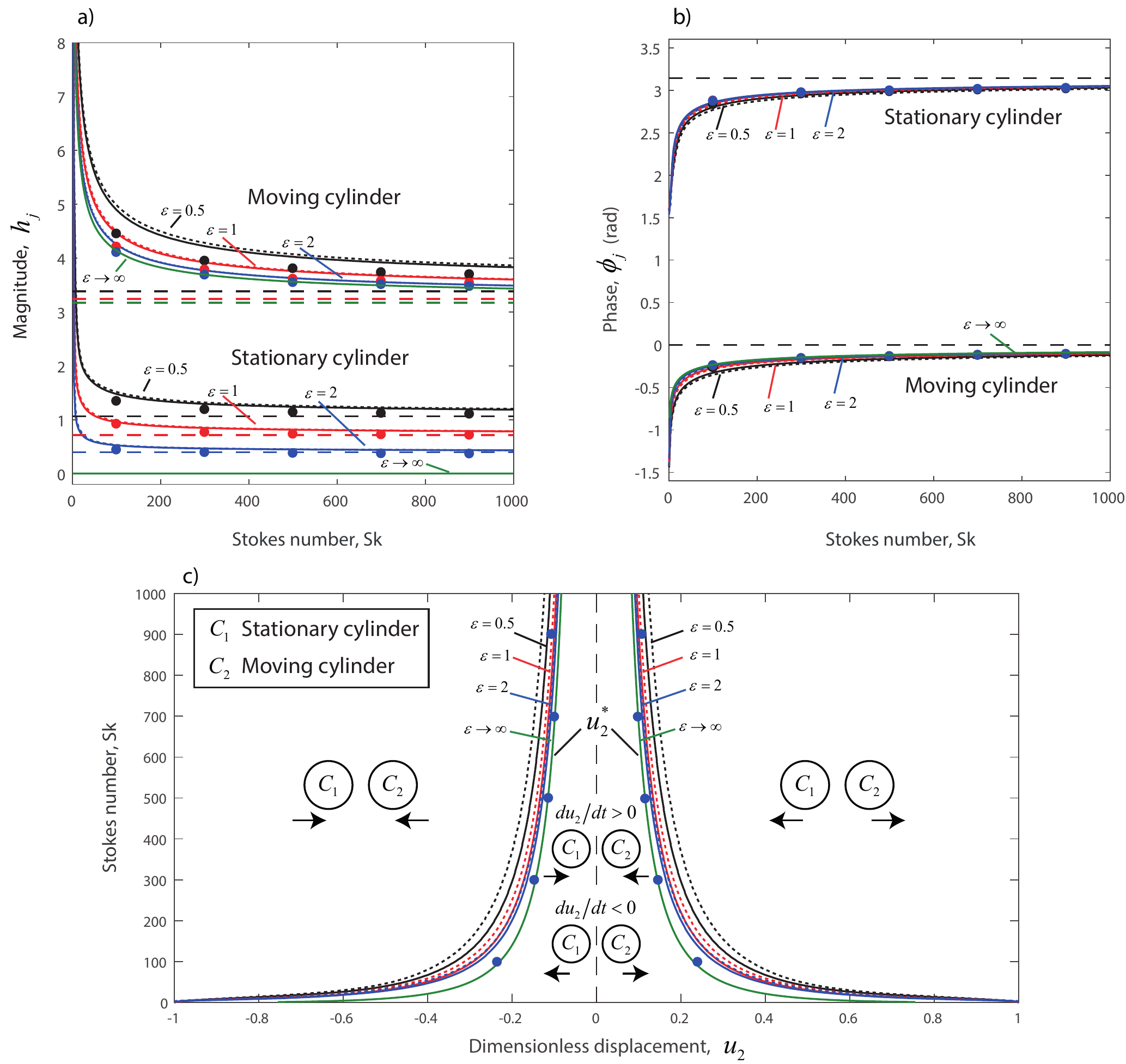}
\caption{Evolutions of a) the magnitude $h_j=|{f_{jx}/u_{2x}}|$ and b) the phase $\phi_j= {\rm{angle}}\left({f_{jx}/u_{2x}}\right)$ versus the Stokes number $Sk$. On c), the arrows show the direction of the fluid forces depending on $Sk$, $\varepsilon$ and the dimensionless displacement of ${\mathcal{C}}_2$. The solid lines refer to the least squares approximation and the dotted lines refer to the collocation approximation. The dimensionless separation distance is $\varepsilon=0.5$ (black color), $\varepsilon=1$ (red color), $\varepsilon=2$ (blue color) and $\varepsilon\rightarrow\infty$  (green color). On a) and b), the horizontal dashed lines are the asymptotic limits as $Sk \rightarrow\infty$. Closed circles correpond to numerical predictions. On b) and c) the black, blue and red circles are indistinguishable because superimposed.}\label{Gain_phase}
\end{center}
\end{figure}

\subsection{Fluid added coefficients}

We now proceed with analyzing the evolutions of the fluid added coefficients $m_{self}$, $c_{self}$, $m_{cross}$ and $c_{cross}$ entering in the computation of the fluid forces.
\\
\\
The evolutions of $m_{self}$ and $c_{self}$ are depicted in {\textcolor{black}{Figures}} \ref{Fig_Added_coefficients} a) and b). We observe that $m_{self}$ and $c_{self}$ diverge to infinity as $Sk\rightarrow 0$ and decrease to $m_{self}\rightarrow m_{self}^{POT}$ and $c_{self}\rightarrow 0$ as $Sk\rightarrow \infty$ (inviscid fluid). The log-log plots in the insets of {\textcolor{black}{Figures}} \ref{Fig_Added_coefficients} a) and b) indicate that 
\begin{equation}\label{eq_scaling_self}
m_{self}  = m_{self}^{Pot}  + O\left( {Sk^{ - {1 \mathord{\left/
 {\vphantom {1 2}} \right.
 \kern-\nulldelimiterspace} 2}} } \right) \;\; {\rm{and}} \;\;
c_{self}  = O\left( {Sk^{ - {1 \mathord{\left/
 {\vphantom {1 2}} \right.
 \kern-\nulldelimiterspace} 2}} } \right) \;\; {\rm{as}} \;\; Sk\rightarrow\infty.
\end{equation} 
In addition to the dependence on the Stokes number, $m_{self}$ and $c_{self}$ are also sensitive to the confinement. The two coefficients are maximum for the small values of $\varepsilon$ (strong confinement) and decrease to $m_{self}\rightarrow m_{self}^{ISO}$ and $c_{self}\rightarrow c_{self}^{ISO}$ as $\varepsilon \rightarrow \infty$ (isolated cylinders). As both $Sk$ and $\varepsilon$ tend to infinity, we recover the classical results for an isolated cylinder in a perfect fluid, $m_{self}\rightarrow 1$ and $c_{self}\rightarrow 0$. 
\\
\\
The evolutions of $m_{cross}$ and $c_{cross}$ are depicted in {\textcolor{black}{Figures}} \ref{Fig_Added_coefficients} c) and d).
We observe that $m_{cross}$ is negative and converges to $m_{cross}\rightarrow 0$ as $Sk\rightarrow 0$. As $Sk$ increases, $m_{cross}$ first decreases, then hits a minimum, and finally increases to $m_{cross}\rightarrow m_{cross}^{POT}$ as $Sk\rightarrow \infty$. 
We hypothesize that the non-monotic variations of $m_{cross}$ are related to an antagonist competition between the viscous and the confinement effects. The term $c_{cross}$ is also negative, diverges to $c_{cross}\rightarrow -\infty$ as $Sk\rightarrow 0$ and increases to $c_{cross}\rightarrow 0$ as $Sk \rightarrow \infty$. The log-log plots in the insets of {\textcolor{black}{Figures}} \ref{Fig_Added_coefficients} c) and d) indicate that 
\begin{equation}\label{eq_scaling_cross}
m_{cross}  = m_{cross}^{Pot}  + O\left( {Sk^{ - {1 \mathord{\left/
 {\vphantom {1 2}} \right.
 \kern-\nulldelimiterspace} 2}} } \right) \;\; {\rm{and}} \;\;
c_{cross}  = O\left( {Sk^{ - {1 \mathord{\left/
 {\vphantom {1 2}} \right.
 \kern-\nulldelimiterspace} 2}} } \right) \;\; {\rm{as}} \;\; Sk\rightarrow\infty.
\end{equation}
The coefficients $m_{cross}$ and $c_{cross}$ are also sensitive to the confinement: they are minimum for the small values of $\varepsilon$ (strong confinement) and increase to $m_{cross} \rightarrow 0$ and $c_{cross} \rightarrow 0$ as $\varepsilon \rightarrow \infty$ (isolated cylinders). In such a case, and as expected, there is no fluid force acting on the stationary cylinder.

Here again, the theoretical predictions for the fluid added coefficients are successfully corroborated by the numerical simulations, in the sense that similar variations are recovered. However, we note that both approaches do not exactly exhibit the same sensitivity to the confinement effect, leading to some deviations in the predictions, in particular concerning the self added coefficients at low Stokes numbers. We discuss the possible origins of these deviations in the next section.

\begin{figure}[H]
\begin{center}
\includegraphics[width=1\textwidth]{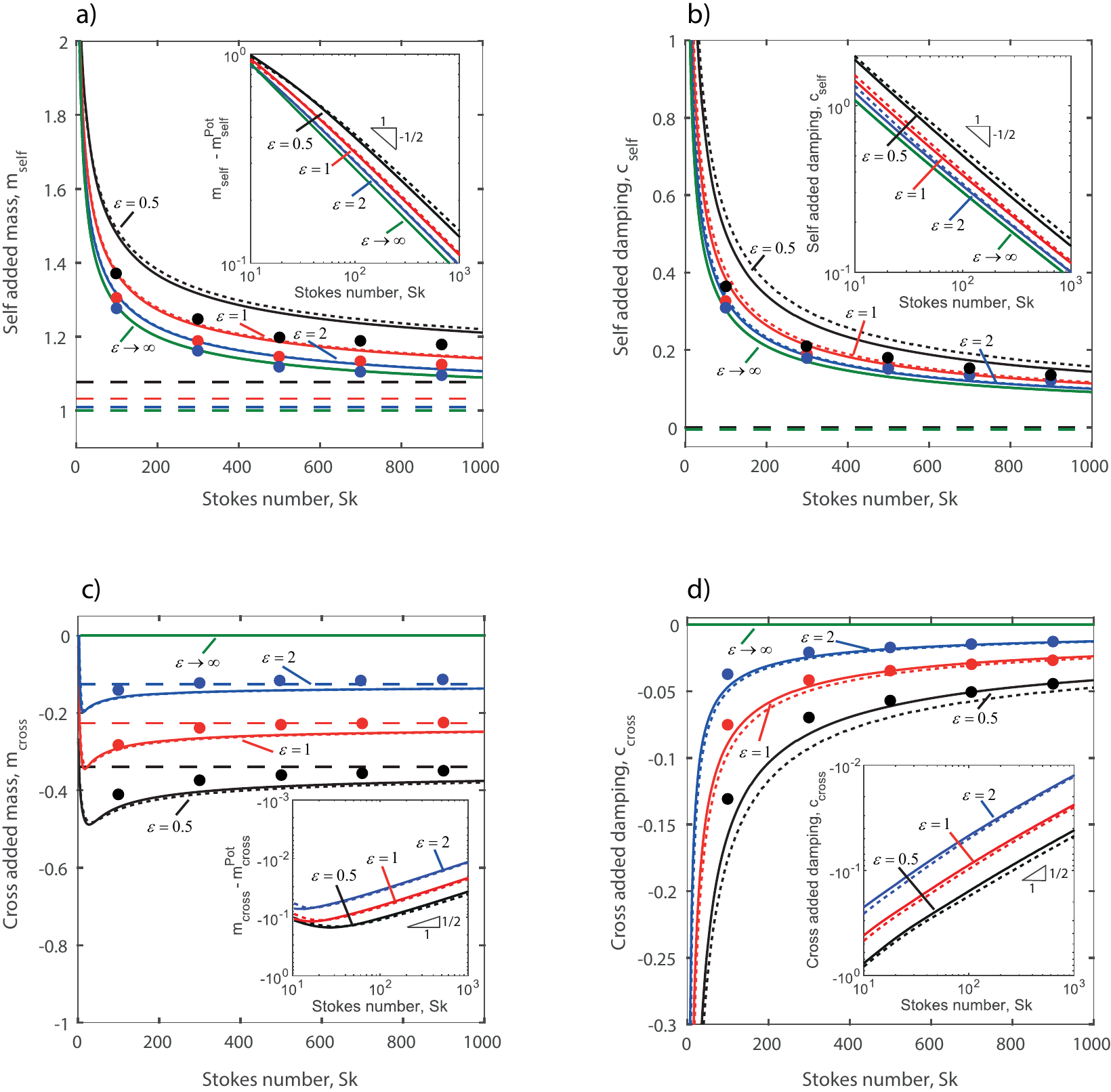}
\caption{Evolutions of the fluid added coefficients versus the Stokes number, $Sk$. The solid lines refer to the least squares approximation and the dotted lines refer to the collocation approximation. The dimensionless separation distance is $\varepsilon=0.5$ (black color), $\varepsilon=1$ (red color), $\varepsilon=2$ (blue color) and $\varepsilon\rightarrow\infty$  (green color). The horizontal dashed lines are the asymptotic limits \eqref{EqAddedMass_SkInf} as $Sk \rightarrow\infty$. Closed circles correpond to numerical predictions.}\label{Fig_Added_coefficients}
\end{center}
\end{figure}

\subsection{Discussion on numerics versus theory}\label{sec:num_vs_theory}

The {\textcolor{black}{Figure}} \ref{Fig_Added_coefficients} shows that the simulations tend to underestimate $m_{self}$ and $c_{self}$, and surestimate $m_{cross}$ and $c_{cross}$. To quantify this deviation, we introduce the quantity $\iota$, defined as the relative distance between the numerical and the theoretical predictions of some quantity $Q$ : $\iota=|Q_{\rm{num.}}-Q_{\rm{th.}}|/|Q_{\rm{num.}}|$. The {\textcolor{black}{Figure}} \ref{Fig_Deviation} and the tables in appendix \ref{Appendix_Tables} show that $\iota$ is maximum for the small values of $Sk$ and $\varepsilon$. We attribute this deviation to the fact that the theoretical approach is based on an approximation (least squares or collocation method) which loses its accurary when $Sk$ and $\varepsilon$ become small, as shown in the study of the residuals in {\textcolor{black}{Figure}} \ref{Fig2} b). Also, the numerical simulation, which is based on a penalization method, hardly makes the difference between the solid and the fluid domains for the low values of $Sk$. {\textcolor{black}{Finally, we shall note that the theoretical approach is fully linear since the convective term $KC{{\left({\bf{{v}}^*} \cdot \nabla\right) {\bf{{v}}^*} }}$ of the Navier-Stokes equation \eqref{DimensionlessNavierStokes1a} is neglected. In the numerical simulations, the nonlinear convective term is retained through a small but nonzero Keulegan-Carpenter number $KC=10^{-2}$. This difference might slightly affect the deviation between the theoretical and numerical results.}}  
In any case, the relative deviation for $m_{self}$ (resp. $m_{cross}$) is always smaller than $\iota \le 
10\%$ (resp. $\iota<20\%$). The deviation for the damping coefficients $c_{self}$ and $c_{cross}$ is more pronounced, with $\iota \le 50\%$ and $\iota \le 35\%$, respectively. Note that the maximum deviations are observed for $Sk\in\left[0,400\right], \varepsilon<1$, and are less important when using the least squares method. Even if the approximations of the theoretical and numerical approaches can be invoked, the slope steepness of the damping coefficients also contributes to the enhancement of the relative deviation in such a range of $Sk$ and $\varepsilon$. It follows that both approaches yield similar trends, bringing out the same behavior of the fluid coefficients, despite some deviations in the particular case of a very viscous fluid (low $Sk$) in a confined environnement ($\varepsilon<1$). 

\begin{figure}[H]
\begin{center}
\includegraphics[width=1\textwidth]{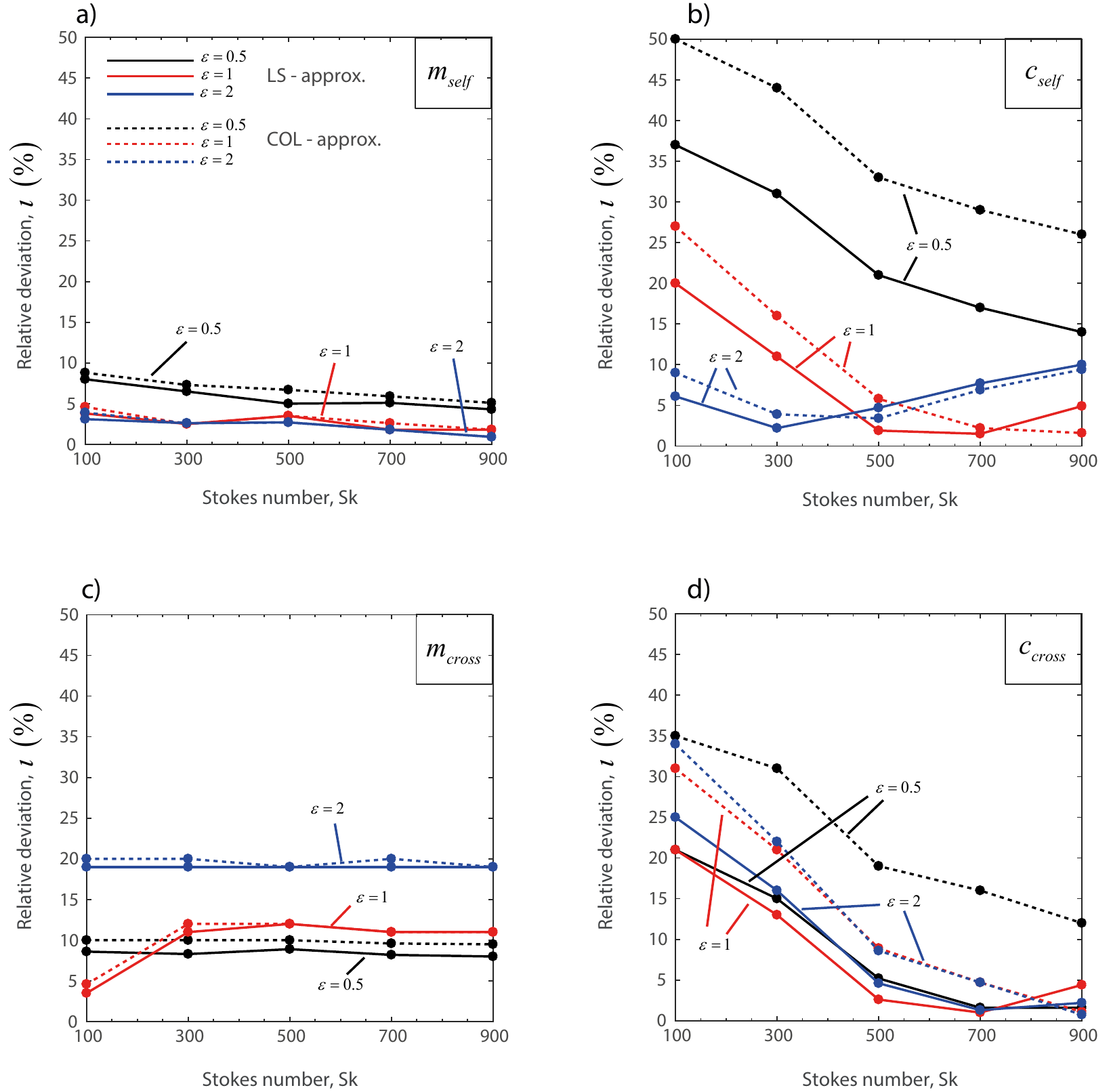}
\caption{Evolutions of the relative deviation, $\iota$, versus the Stokes number, $Sk$. The solid lines refer to the least squares approximation while the dotted lines refer to the collocation approximation. The dimensionless separation distance is $\varepsilon=0.5$ (black color), $\varepsilon=1$ (red color), $\varepsilon=2$ (blue color).}\label{Fig_Deviation}
\end{center}
\end{figure}

\section{{\textcolor{black}{Conclusions}}}\label{Sec:Conclusion}

We have considered the problem of the small oscillations of two cylinders immersed in a viscous fluid initially at rest. A theoretical approach based on an Helmholtz decomposition of the fluid velocity vector and a bipolar coordinate system has been carried out to estimate the fluid forces acting on the two cylinders. 
In addition to this new theoretical work, we also have developed a numerical approach based on a pseudo-penalization method. Such a numerical method has been shown particularly efficient in solving fluid-structure interaction problems, in particular for moderate or high Stokes numbers.  

We studied the case in which one cylinder is {\textcolor{black}{stationary}} while the other one is imposed {\textcolor{black}{a harmonic}} motion. We show that the amplitude, the phase and the direction of the fluid forces are sensitive to the Stokes number and the separation distance between the cylinders. The two forces are in phase opposition and their amplitude decreases to the inviscid limits as $Sk$ increases. The effect of viscosity is to add to the ideal fluid added coefficients a correction term which scales as $Sk^{-1/2}$. When the separation distance increases, the fluid coefficients converge to the limits of an isolated cylinder derived by Stokes \cite{Stokes1850}. The theoretical predictions are successfully corroborated by the numerical simulations, in the sense that similar trends are recovered, despite some deviations for low 
$Sk$ and $\varepsilon$.      
 
As an improvement to our previous work on ideal fluids \cite{Lagrange2018}, the new theoretical approach carried out in the present article is able to capture the effects of viscosity on the fluid forces. It offers a simple and flexible alternative to the fastidious and hardly tractable approach developed by \cite{Chen1975b}. To our knowledge, this is also the first time that the pseudo-penalization method is presented in the context of relatively small Stokes numbers. As such, the present work should foster further developements of this easy to implement numerical method, to tackle complex fluid-structure interaction problems.

%\section*{Acknowledgment}
%The authors would like to thank J. Antunes for his valuable comments and suggestions to improve the manuscript.

\appendix

\textcolor{black}{\section{Evolutions of $m_{self}^{(j)POT}$ and $m_{cross}^{POT}$}}\label{Appendix_Inviscid}
In this appendix, we study the variations of the fluid added coefficients $m_{self}^{(j)POT}$ and $m_{cross}^{POT}$, given by \eqref{EqAddedMass_SkInf}.  
{\textcolor{black}{In Figure \ref{Fig_AddedMass_SkInf_JFS2017}, we show their evolution with the dimensionless separation distance $\varepsilon$, considering two equal size cylinders, i.e. $r=1$, for which $m_{self}^{(1)POT}=m_{self}^{(2)POT}=m_{self}^{POT}$. We observe that $m_{self}^{POT}$ (resp. $m_{cross}^{POT}$) decreases (resp. increases) monotonically with the dimensionless separation distance. When the cylinders are in close proximity, i.e. ${\varepsilon} \to 0$, the confinement is maximum and the added coefficients become unbounded, as expected. When the two cylinders are far apart, i.e. ${\varepsilon} \to \infty $, they both behave like an isolated cylinder in an infinite fluid domain, $m_{self}^{POT} \to 1 $ and $m_{cross}^{POT} \to 0 $. To validate our observations, we have reported in Figure \ref{Fig_AddedMass_SkInf_JFS2017} the predictions of the literature \cite{Mazur1970,Landweber1991}. Unlike the current method, \cite{Mazur1970} used a conformal mapping method to solve the potential problem and extracted the potential added mass coefficients from the kinetic energy of the fluid. On his side, \cite{Landweber1991} extended the method of images by \cite{Hicks1879,Herman1887} and extracted the added mass coefficients from the fluid force acting on the cylinders. We obtain an excellent agreement with those authors, thereby validating our prediction for $m_{self}^{POT}$ and $m_{cross}^{POT}$ for $r=1$.}}

In {\textcolor{black}{Figure}} \ref{Fig_AddedMass_SkInf}, we show that $m_{self}^{(j)POT}$ (resp. $m_{cross}^{POT}$) increases (resp. decreases) with the radius ratio $r$ while it decreases (resp. increases) with the dimensionless separation distance $\varepsilon$. When $r\rightarrow 0$, the cylinder ${\mathcal{C}}_1$ transforms to a point and the system is equivalent to an isolated cylinder ${\mathcal{C}}_2$, leading to the classical result $m_{self}^{(2)POT}\rightarrow 1$.    
On the other hand, when $r\rightarrow\infty$, the cylinder ${\mathcal{C}}_1$ transforms to an infinite plane and the system is equivalent to a cylinder ${\mathcal{C}}_2$ near a wall. In such a case, we obtain
\begin{subequations}\label{m_WALL}
\begin{align}
m_{self}^{\left( 1 \right)POT}  &\to \infty,\\
m_{self}^{\left( 2 \right)POT}  &\to m_{self}^{WALL}=-4\sum\limits_{n = 1}^\infty {\frac {n\varepsilon  \left( 2+\varepsilon \right)  \left( {4}^{n}+{16}
^{n} \left( 2 \varepsilon+2 \sqrt {\varepsilon  \left( 2+\varepsilon
 \right) }+2 \right) ^{-2\,n} \right) }{- \left( 2 \varepsilon+2 \sqrt 
{\varepsilon  \left( 2+\varepsilon \right) }+2 \right) ^{2 n}+{4}^{n}}},\\
m_{cross}^{POT}  &\to m_{cross}^{WALL}=8\sum\limits_{n = 1}^\infty {\frac { {4}^{n} n \varepsilon\left( 2+\varepsilon \right) }{{4}^{n}-
 \left( 2 \sqrt {\varepsilon \left( 2+\varepsilon \right) }+2 \varepsilon+2
 \right) ^{2 n}}}.
\end{align}
\end{subequations}
Values of $m_{self}^{WALL}$ are presented in {\textcolor{black}{Figure}} \ref{Fig_AddedMass_rInf}, showing a perfect agreement with the predictions of \cite{Mazur1966} and \cite{Chen1987}. 

\begin{figure}[H]
\begin{center}
\includegraphics[width=1\textwidth]{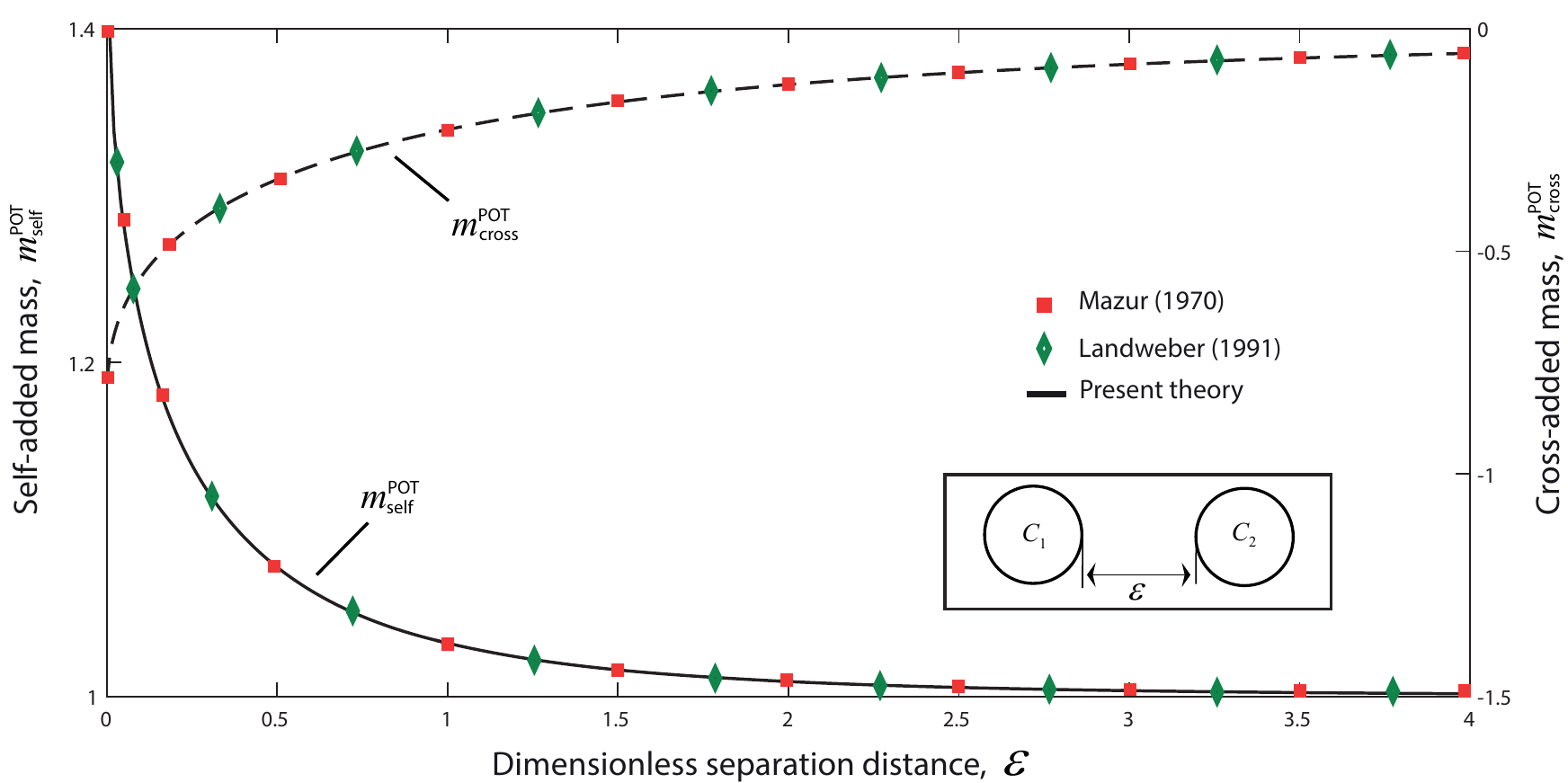}
\caption{Fluid added coefficients $m_{self}^{POT}$ and $m_{cross}^{POT}$, given by Eq. \eqref{EqAddedMass_SkInf}, versus the dimensionless separation distance $\varepsilon$. The radius ratio is $r=1$.}\label{Fig_AddedMass_SkInf_JFS2017}
\end{center}
\end{figure}

\begin{figure}[H]
\begin{center}
\includegraphics[width=1\textwidth]{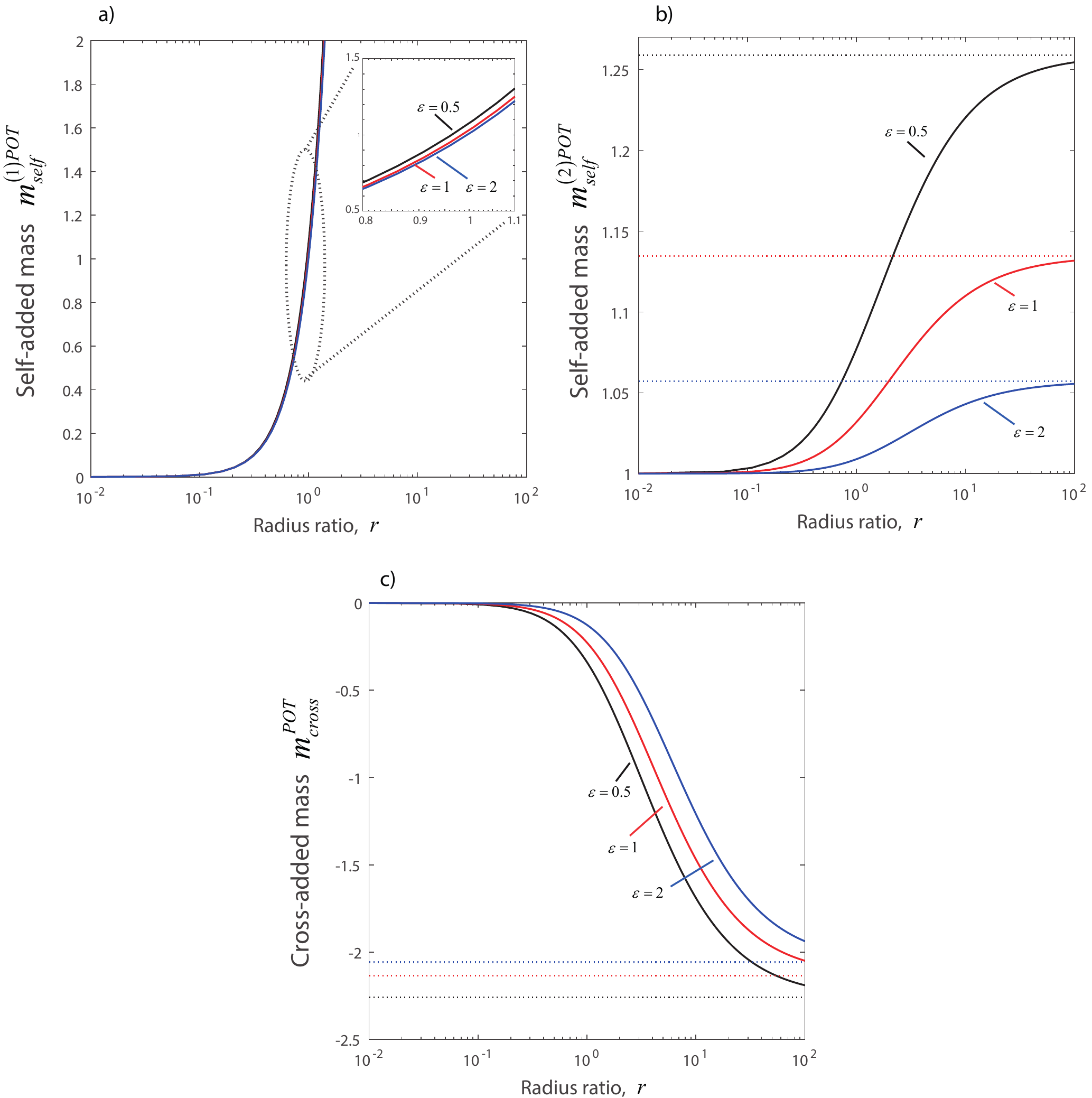}
\caption{Fluid added coefficients, given by Eq. \eqref{EqAddedMass_SkInf}, versus the radius ratio $r$. Evolution of a) $m_{self}^{(1)POT}$, b) $m_{self}^{(2)POT}$ and c) $m_{cross}^{POT}$. The horizontal dotted lines on b) and c) show the limits as $r\rightarrow\infty$, see {\textcolor{black}{Eq.}} \eqref{m_WALL}. The dimensionless separation distance is $\varepsilon=0.5$ (black color), $\varepsilon=1$ (red color), $\varepsilon=2$ (blue color).}\label{Fig_AddedMass_SkInf}
\end{center}
\end{figure}

\begin{figure}[H]
\begin{center}
\includegraphics[width=0.5\textwidth]{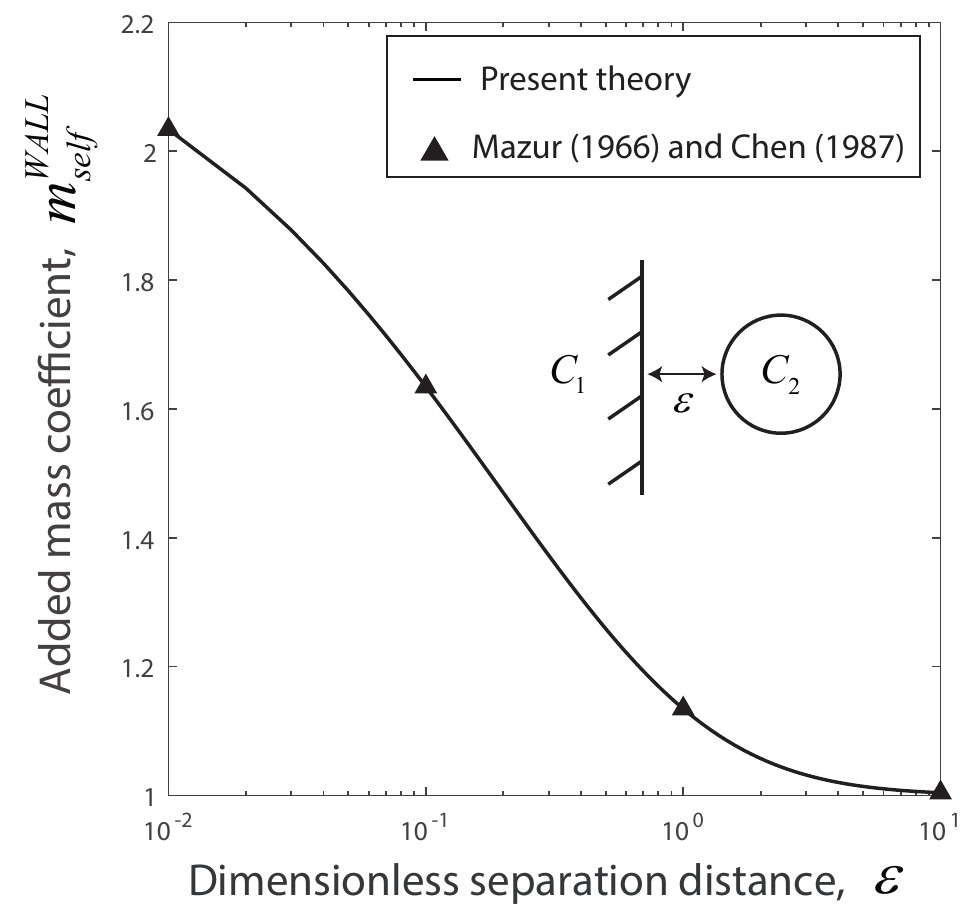}
\caption{Added mass coefficient $m_{self}^{WALL}$, given by \eqref{m_WALL} for a cylinder vibrating near a wall.}\label{Fig_AddedMass_rInf}
\end{center}
\end{figure}

\textcolor{black}{\section{Functions $\varphi _n^{(j)}$ and $A _n^{(j)}$}}\label{Appendix_Constants_Integration}
The functions $\varphi _n^{(j)}$ and $A _n^{(j)}$ appearing in \eqref{Fluid_functions_bipolar} are determined from the boundary conditions \eqref{BC1_bipolar}, \eqref{BC2_bipolar}. It yields a linear system of equations, whose solution is
\begin{equation}
\left[ \begin {array}{c} \varphi _n^{(1)} \left( \xi_{{1}},\xi
_{{2}},l \right) \\\noalign{\medskip}\varphi _n^{(2)} \left( 
\xi_{{1}},\xi_{{2}},l \right) \\\noalign{\medskip}{ A _n^{(1)} } \left( \xi
_{{1}},\xi_{{2}},l \right) \\\noalign{\medskip}{ A _n^{(2)} } \left( \xi_{{
1}},\xi_{{2}},l \right) \end {array} \right] = \left[ M_{{n}} \left( 
\xi_{{1}},\xi_{{2}},l \right)  \right] ^{-1} \left[ 
\begin {array}{c} -2\,na{{\rm e}^{-n \left| \xi_{{1}} \right| }}{\it 
\sign} \left( \xi_{{1}} \right) \\\noalign{\medskip}-2\,na{{\rm e}^{-
n \left| \xi_{{1}} \right| }}\\\noalign{\medskip}0\\\noalign{\medskip}0
\end {array} \right], 
\end{equation}
with
\begin{equation}
\left[M_n\left(\xi_1,\xi_2,l\right)\right] = \left[ \begin {array}
{cccc} -n\cosh \left( n\xi_{{1}} \right) &-n\sinh \left( n\xi_{{1}}
 \right) &\sinh \left( l\xi_{{1}} \right) l&\cosh \left( l\xi_{{1}}
 \right) l\\\noalign{\medskip}n\sinh \left( n\xi_{{1}} \right) &n\cosh
 \left( n\xi_{{1}} \right) &-n\cosh \left( l\xi_{{1}} \right) &-n\sinh
 \left( l\xi_{{1}} \right) \\\noalign{\medskip}-n\cosh \left( n\xi_{{2
}} \right) &-n\sinh \left( n\xi_{{2}} \right) &\sinh \left( l\xi_{{2}}
 \right) l&\cosh \left( l\xi_{{2}} \right) l\\\noalign{\medskip}n\sinh
 \left( n\xi_{{2}} \right) &n\cosh \left( n\xi_{{2}} \right) &-n\cosh
 \left( l\xi_{{2}} \right) &-n\sinh \left( l\xi_{{2}} \right) 
\end {array} \right].
\end{equation}

\textcolor{black}{\section{Effect of the mesh size, time step and computational domain size on the fluid added coefficients}\label{Appendix_Effect_mesh_time}}
{\textcolor{black}{In this appendix, we report the numerical values of the fluid added coefficients obtained with different mesh sizes, time steps and computational domain sizes. We have considered the case of two equal size cylinders, i.e. $r=1$, a dimensionless separation distance $\varepsilon=0.5$ and a Stokes number $Sk=300$. In Tables \ref{Table_mesh_effect}, \ref{Table_time_step_effect} and \ref{Table_domain_effect}, we clearly show that refining the mesh size ($x\times y\rightarrow 2x\times 2y$), reducing the time step ($\delta t\rightarrow \delta t/2$) or increasing the computational domain size ($L_x\times L_y \rightarrow 2L_x\times 2L_y$), has no significant effect on the fluid coefficients. From this observation, we conclude that the results shown in the main core of the manuscript (obtained for $x\times y=3060\times 1850$, $\delta t=5\times 10^{-3}$ and $L_x\times L_y=20\times 17$) are actually very poorly sensitive to $x\times y$, $\delta t$ and $L_x\times L_y$.}}  

\begin{table}[H]
\begin{center}
\begin{tabular}{ccccc}
mesh size & $m_{self}$ & $c_{self}$ & $m_{cross}$ & $c_{cross}$\\
\hline
$x\times y$ & 1.24 & 0.208 & -0.372 & -0.0706 \\
$2x\times 2y$ & 1.257 & 0.201 & -0.374 & -0.0681 \\
\end{tabular}
\caption{Effect of the mesh size on the fluid added coefficients. The mesh size used in Section \ref{Sec_results} is $x\times y=3060\times 1850$ and the time step is $\delta t= 5\times 10^{-3}$. The time step used for the mesh size $2x\times 2y$ is $\delta t= 1.25\times 10^{-3}$. The dimensionless separation distance is $\varepsilon=0.5$ and the Stokes number is $Sk=300$.} \label{Table_mesh_effect} 
\end{center}
\end{table}
				
\begin{table}[H]
\begin{center}
\begin{tabular}{ccccc}
Time step & $m_{self}$ & $c_{self}$ & $m_{cross}$ & $c_{cross}$\\
\hline
$\delta t$ & 1.24 & 0.208 & -0.372 & -0.0706 \\
$\delta t/2$ & 1.259 & 0.203 & -0.375 & -0.069 \\
\end{tabular}
\caption{Effect of the time step on the fluid added coefficients. The time step used in Section \ref{Sec_results} is $\delta t= 5\times 10^{-3}$. The mesh size is $x\times y=3060\times 1850$. The dimensionless separation distance is $\varepsilon=0.5$ and the Stokes number is $Sk=300$.} \label{Table_time_step_effect} 
\end{center}
\end{table}

\begin{table}[H]
\begin{center}
\begin{tabular}{ccccc}
Domain size & $m_{self}$ & $c_{self}$ & $m_{cross}$ & $c_{cross}$\\
\hline
$L_x \times L_y$ & 1.24 & 0.208 & -0.372 & -0.0706 \\
$2 L_x \times 2 L_y$ & 1.227 & 0.206 & -0.386 & -0.0727 \\
\end{tabular}
\caption{Effect of the computational domain size on the fluid added coefficients. The computational domain size used in Section \ref{Sec_results} is $L_x \times L_y=20\times 17$. The mesh size is $x\times y=3060\times 1850$ on $L_x \times L_y$, and nonuniform on the rest of the domain size. The time step is $\delta t= 5\times 10^{-3}$. The dimensionless separation distance is $\varepsilon=0.5$ and the Stokes number is $Sk=300$.}\label{Table_domain_effect}  
\end{center}
\end{table}
\textcolor{black}{\section{Tables of comparison numerics versus theory}\label{Appendix_Tables}}

In this appendix, we report the theoretical and numerical values of the fluid added coefficients $m_{self}$, $c_{self}$, $m_{cross}$ and $c_{cross}$, for $\varepsilon=0.5$ (table  \ref{Comparison_theory_numerics_eps05}), $\varepsilon=1$ (table \ref{Comparison_theory_numerics_eps1}) and $\varepsilon=2$ (table \ref{Comparison_theory_numerics_eps2}). The numerical values correspond to the closed symbols shown in {\textcolor{black}{Figure}} \ref{Fig_Added_coefficients}. The relative deviation $\iota$ is also reported in the tables.
  
\begin{table}[H]
\begin{center}
\includegraphics[width=1\textwidth]{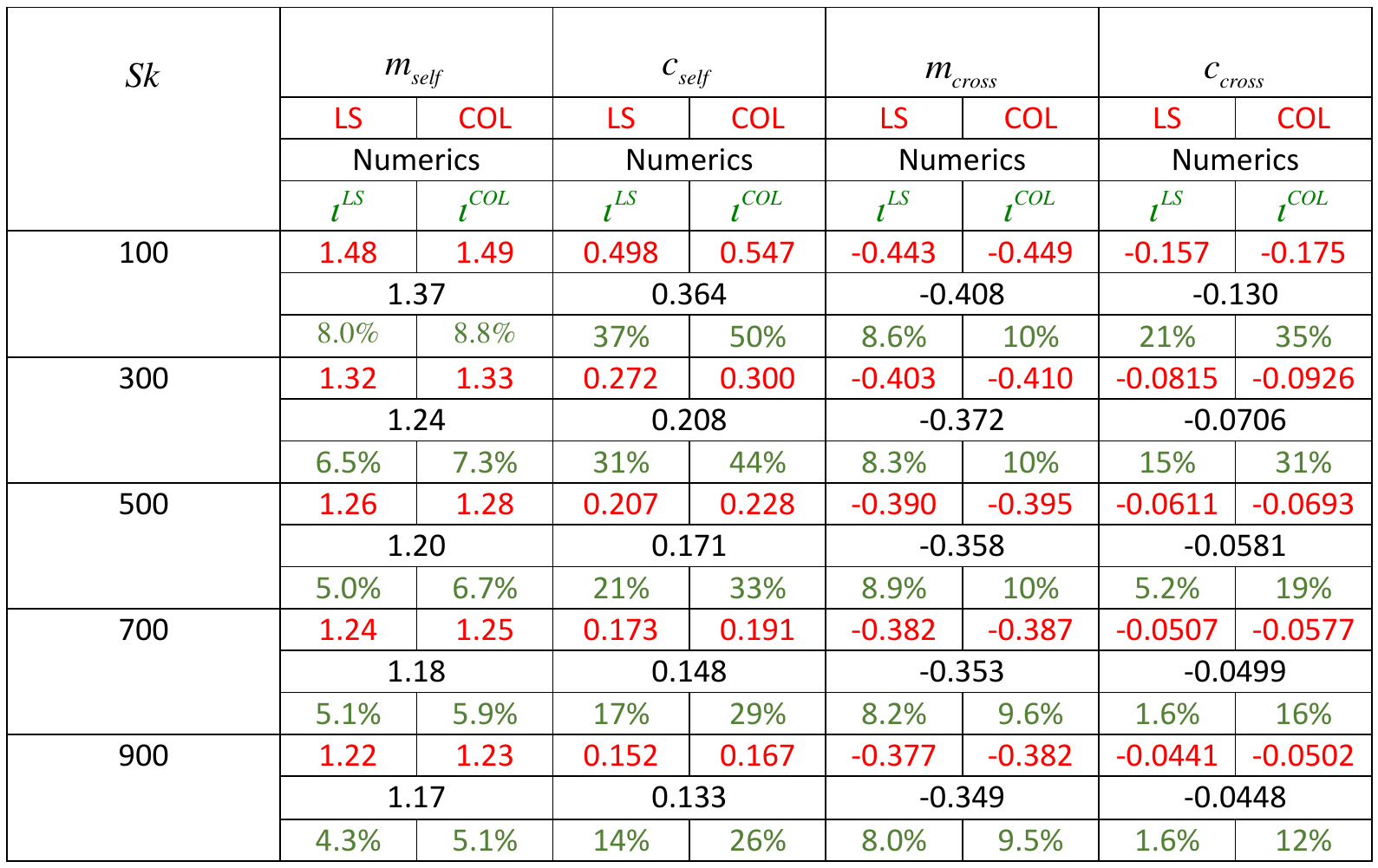}
\caption{Table of the fluid added coefficients and the relative deviation, $\iota$. The notations LS and COL refer to the Least Squares and Collocation methods. The dimensionless separation distance is $\varepsilon=0.5$.}\label{Comparison_theory_numerics_eps05}
\end{center}
\end{table}

\begin{table}[H]
\begin{center}
\includegraphics[width=1\textwidth]{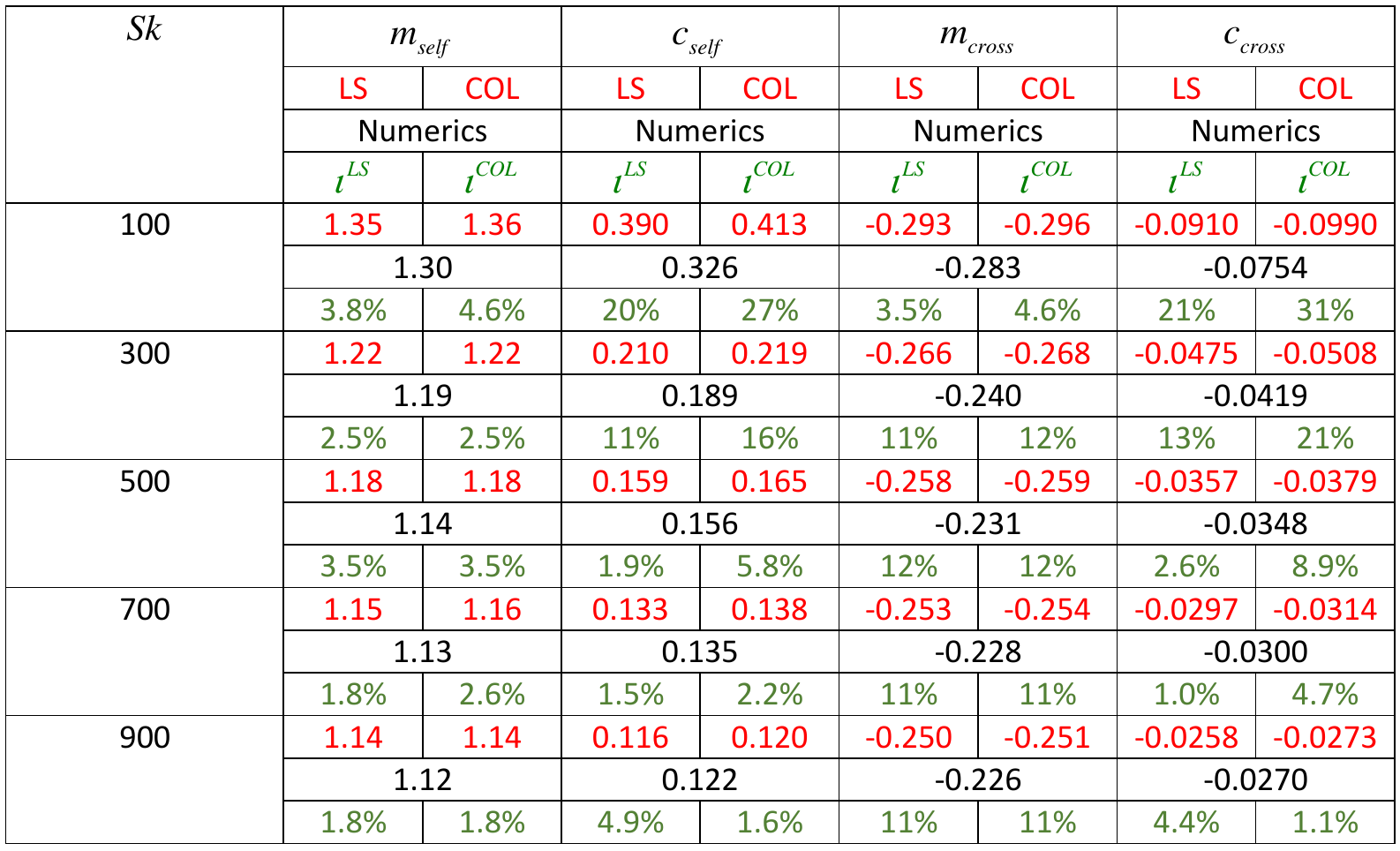}
\caption{Table of the fluid added coefficients and the relative deviation, $\iota$. The notations LS and COL refer to the Least Squares and Collocation methods. The dimensionless separation distance is $\varepsilon=1$.}\label{Comparison_theory_numerics_eps1}
\end{center}
\end{table}

\begin{table}[H]
\begin{center}
\includegraphics[width=1\textwidth]{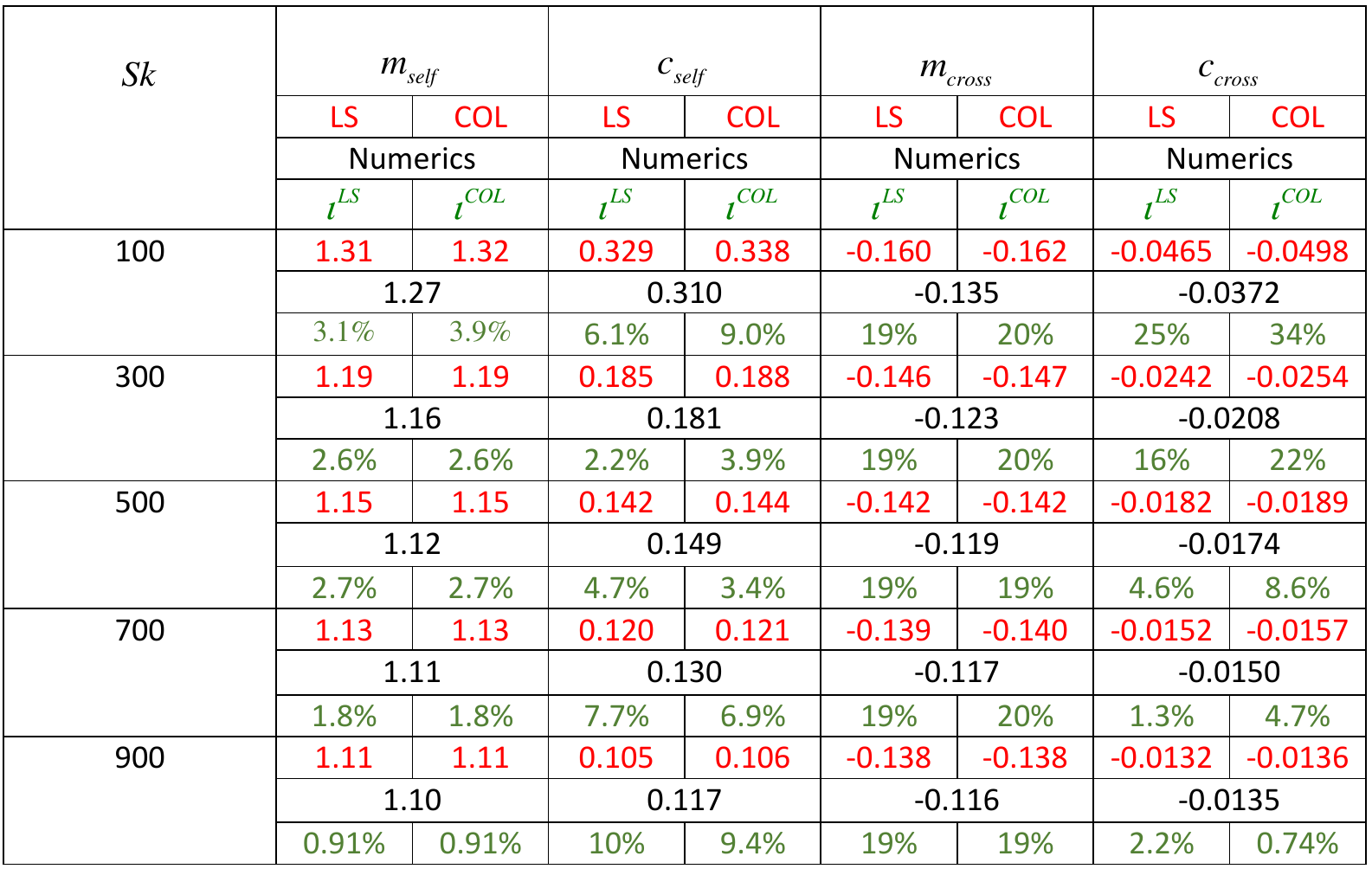}
\caption{Table of the fluid added coefficients and the relative deviation, $\iota$. The notations LS and COL refer to the Least Squares and Collocation methods. The dimensionless separation distance is $\varepsilon=2$.}\label{Comparison_theory_numerics_eps2}
\end{center}
\end{table}

\bibliographystyle{elsarticle-num}
\bibliography{biblio}

\end{document}